\documentclass[twocolumn]{aastex61}
\begin{document}

%\title{Direct Imaging Constraints on the (Non-)Existence of Protoplanets Around LkCa 15 from SCExAO/CHARIS and Keck/NIRC2}
%\title{Evidence for the Non-Existence of Protoplanets Around LkCa 15\\  from SCExAO/CHARIS and Keck/NIRC2}
\title{No Clear, Direct Evidence for Multiple Protoplanets Orbiting LkCa 15: 
 \\LkCa 15 \lowercase{bcd} are Likely Inner Disk Signals}
 % \\LkCa 15 \lowercase{bcd} are Likely (Contaminated By) Inner Disk Signals}
\correspondingauthor{Thayne Currie}
\email{thayne.m.currie@nasa.gov,currie@naoj.org}
\author{Thayne Currie}
\affiliation{NASA-Ames Research Center, Moffett Blvd., Moffett Field, CA, USA}
\affiliation{Subaru Telescope, National Astronomical Observatory of Japan, 
650 North A`oh$\bar{o}$k$\bar{u}$ Place, Hilo, HI  96720, USA}
\affiliation{Eureka Scientific, 2452 Delmer Street Suite 100, Oakland, CA, USA}
\author{Christian Marois}
\affiliation{National Research Council of Canada Herzberg, 5071 West Saanich Rd, Victoria, BC, V9E 2E7, Canada}
\affiliation{University of Victoria, 3800 Finnerty Rd, Victoria, BC, V8P 5C2, Canada}
\author{Lucas Cieza}
\affiliation{N\'ucleo de Astronom\'ia, Facultad de Ingenier\'ia y Ciencias, Universidad Diego Portales, Av Ej\'ercito 441, Santiago, Chile}
%\affiliation{Nucleo de Astronomia, University of Diego Portales, Santiago, Chile}
\author{Gijs D. Mulders}
\affil{"Department of the Geophysical Sciences, The University of Chicago, Chicago, IL 60637, USA}
%\author{Michiel Min}
%\affiliation{SRON Netherlands Institute for Space Research, Sorbonnelaan 2, 3584 CA, Utrecht, The Netherlands}
%\affiliation{Astronomical institute Anton Pannekoek, University of Amsterdam, Science Park 904, 1098 XH, Amsterdam, The Netherlands}
\author{Kellen Lawson}
\affiliation{Homer L. Dodge Department of Physics, University of Oklahoma, Norman, OK 73071, USA}
\author{Claudio Caceres}
\affiliation{Departamento de Ciencias Fisicas, Facultad de Ciencias Exactas, Universidad Andres Bello. Av. Fernandez Concha 700, Las Condes, Santiago, Chile}
\affiliation{N\'ucleo Milenio Formaci\'on Planetaria - NPF, Universidad de Valpara\'iso, Av. Gran Breta\~na 1111, Valpara\'iso, Chile}
\author{Dary Rodriguez-Ruiz}
\affiliation{Department of Physics and Astronomy, Rochester Institute of Technology, Rochester, NY, USA}
%\affiliation{Department of Physics, University of Diego Portales, Santiago, Chile}
\author{John Wisniewski}
\affiliation{Homer L. Dodge Department of Physics, University of Oklahoma, Norman, OK 73071, USA}
\author{Olivier Guyon}
\affiliation{Subaru Telescope, National Astronomical Observatory of Japan, 
650 North A`oh$\bar{o}$k$\bar{u}$ Place, Hilo, HI  96720, USA}
\affil{Steward Observatory, University of Arizona, Tucson, AZ 85721, USA}
\affil{College of Optical Sciences, University of Arizona, Tucson, AZ 85721, USA}
\affil{Astrobiology Center of NINS, 2-21-1, Osawa, Mitaka, Tokyo, 181-8588, Japan}
\author{Timothy D. Brandt}
\affiliation{Department of Physics, University of California, Santa Barbara, Santa Barbara, California, USA}
\author{N. Jeremy Kasdin}
\affiliation{Department of Mechanical Engineering, Princeton University, Princeton, NJ, USA}
\author{Tyler D. Groff}
\affiliation{NASA-Goddard Space Flight Center, Greenbelt, MD, USA}
\author{Julien Lozi}
\affiliation{Subaru Telescope, National Astronomical Observatory of Japan, 
650 North A`oh$\bar{o}$k$\bar{u}$ Place, Hilo, HI  96720, USA}
\author{Jeffrey Chilcote}
\affiliation{Department of Physics, University of Notre Dame, South Bend, IN, USA}
\author{Klaus Hodapp}
\affiliation{Institute for Astronomy, University of Hawaii,
640 North A`oh$\bar{o}$k$\bar{u}$ Place, Hilo, HI 96720, USA}
\author{Nemanja Jovanovic}
\affiliation{Department of Astronomy, California Institute of Technology, 1200 East California Boulevard, Pasadena, CA 91125}
\author{Frantz Martinache}
\affiliation{Universit\'{e} C\^{o}te d'Azur, Observatoire de la C\^{o}te d'Azur, CNRS, Laboratoire Lagrange, France}
\author{Nour Skaf}
\affiliation{Subaru Telescope, National Astronomical Observatory of Japan, 
650 North A`oh$\bar{o}$k$\bar{u}$ Place, Hilo, HI  96720, USA}
\affiliation{Imperial College London, Kensington, London SW7 2AZ, UK}
\author{Wladimir Lyra}
\affiliation{Department of Physics and Astronomy, California State University Northridge, 18111 Nordhoff Street, Northridge CA 91130, USA}
\affiliation{Jet Propulsion Laboratory, California Institute of Technology, 4800 Oak Grove Drive, Pasadena, CA, 91109, USA }
\author{Motohide Tamura}
\affil{Astrobiology Center of NINS, 2-21-1, Osawa, Mitaka, Tokyo, 181-8588, Japan}
\affiliation{Department of Astronomy, Graduate School of Science, The University of Tokyo, 7-3-1, Hongo, Bunkyo-ku, Tokyo, 113-0033, Japan}
\affiliation{National Astronomical Observatory of Japan, 2-21-2, Osawa, Mitaka, Tokyo 181-8588, Japan}
%\affiliation{Institut d'Optique Graduate School Paris Saclay, 2 Av. Augustin Fresnel Palaiseau, 91120 France}
\author{Ruben Asensio-Torres}
\affiliation{Department of Astronomy, Stockholm University, AlbaNova University Center, SE-106 91 Stockholm, Sweden}
\author{Ruobing Dong}
\affiliation{University of Victoria, 3800 Finnerty Rd, Victoria, BC, V8P 5C2, Canada}
\author{Carol Grady}
\affiliation{Eureka Scientific, 2452 Delmer Street Suite 100, Oakland, CA, USA}
\affiliation{NASA-Goddard Space Flight Center, Greenbelt, MD, USA}
\author{Misato Fukagawa}
\affiliation{National Astronomical Observatory of Japan, 2-21-2, Osawa, Mitaka, Tokyo 181-8588, Japan}
%\affiliation{Department of Physics, Nagoya University, Nagoya, Japan}
\author{Derek Hand}
\affiliation{Subaru Telescope, National Astronomical Observatory of Japan, 
650 North A`oh$\bar{o}$k$\bar{u}$ Place, Hilo, HI  96720, USA}
\author{Masahiko Hayashi}
\affiliation{National Astronomical Observatory of Japan, 2-21-2, Osawa, Mitaka, Tokyo 181-8588, Japan}
\author{Thomas Henning}
\affiliation{Max Planck Institut fur Astronomie, Konigstuhl 17, 69117 Heidelberg, Germany}
\author{Tomoyuki Kudo}
\affiliation{Subaru Telescope, National Astronomical Observatory of Japan, 
650 North A`oh$\bar{o}$k$\bar{u}$ Place, Hilo, HI  96720, USA}
\author{Masayuki Kuzuhara}
\affil{Astrobiology Center of NINS, 2-21-1, Osawa, Mitaka, Tokyo, 181-8588, Japan}
%\affiliation{Department of Astronomy, University of Tokyo}
\author{Jungmi Kwon}
\affiliation{ISAS/JAXA, 3-1-1 Yoshinodai, Chuo-ku, Sagamihara, Kanagawa 252-5210, Japan}
\author{Michael W. McElwain}
\affiliation{NASA-Goddard Space Flight Center, Greenbelt, MD, USA}
\author{Taichi Uyama}
\affiliation{Department of Astronomy, Graduate School of Science, The University of Tokyo, 7-3-1, Hongo, Bunkyo-ku, Tokyo, 113-0033, Japan}
\begin{abstract}
Two studies utilizing sparse aperture-masking (SAM) interferometry and $H_{\rm \alpha}$ differential imaging
 have reported multiple jovian companions around the young solar-mass star, LkCa 15 (LkCa 15 bcd): the first claimed direct detection of infant, newly formed planets (``protoplanets").
 We present new near-infrared direct imaging/spectroscopy from the Subaru Coronagraphic Extreme Adaptive Optics (SCExAO) system coupled with the Coronagraphic High Angular Resolution Imaging Spectrograph (CHARIS) integral field spectrograph and multi-epoch thermal infrared imaging from Keck/NIRC2 of LkCa 15 at high Strehl ratios.  These data provide the first direct imaging look at the same wavelengths and in the same locations where previous studies identified the LkCa 15 protoplanets, and thus offer the first decisive test of their existence.  
 \\\\   
 The data do not reveal these planets.   Instead, we resolve extended emission tracing a dust disk with a brightness and location comparable to that claimed for LkCa 15 bcd.   Forward-models attributing this signal to orbiting planets are inconsistent with the combined SCExAO/CHARIS and Keck/NIRC2 data.  An inner disk provides a more compelling explanation for the SAM detections and perhaps also the claimed $H_{\alpha}$ detection of LkCa 15 b.      
\\\\
 We conclude that there is currently no clear, direct evidence for multiple protoplanets orbiting LkCa 15, although the system likely contains at least one unseen jovian companion.   To identify jovian companions around LkCa 15 from future observations, the inner disk should be detected and its effect modeled, removed, and shown to be distinguishable from planets.  Protoplanet candidates identified from similar systems should likewise be clearly distinguished from disk emission through modeling.
\end{abstract}
\section{Introduction}
%\textbf{\textit{STUFF NOT DONE YET: 
%\begin{itemize}
%\item show sequence of subtractions image - simulated point source, image - simulated - extended source (Fig 3, right)
%\item Finish photometry for HD 100546 c
%\item image showing predicted position of HD 100546 c/epoch or simulated pt source/extended source subtraction (Fig 4. right)
%\item Get prediction for HD 100546 c in the near-future
%\end{itemize}}}
Young, 1--10 $Myr$-old jovian protoplanets embedded in disks around newly born stars provide a crucial link between the first stages of planet formation and the properties of directly imaged, fully formed planets orbiting 10--100 $Myr$ old stars \citep[e.g.][]{Marois2008b,Marois2010a}.   LkCa 15, a solar-mass T Tauri star and member of the 1--3 $Myr$ old Taurus--Auriga star-forming region \citep{Kenyon2008}, is a superb laboratory for studying planet formation and searching for protoplanets.    The star is surrounded by an accreting, gas-rich protoplanetary disk with multiple dust components: hot ($T_{\rm eff}$ = 1400 $K$), sub-au scale dust producing broadband near-infrared (NIR) excess
 %from sub-au scale dust 
 and cooler massive outer dust, which are separated by a solar system-scale cavity plausibly created by jovian protoplanets \citep{Espaillat2007, Thalmann2010,Andrews2011,DodsonRobinson2011,DongFung2016,Alencar2018}.   
 %The outer disk may show evidence for a pericenter offset plausibly induced by a protoplanet \citep{Thalmann2010}.

Using sparse aperture masking interferometry \citep[SAM;][]{Tuthill2006} of LkCa 15, \citet{Kraus2012} reported the detection of one protoplanet located within an ostensibly cleared gap in dust emission.
%\citep{Espaillat2007,Thalmann2010}.  
% The signal at these two wavelengths hinted at a complex structure, suggesting the detection of a protoplanet surrounded by infalling material.  
Also using SAM, \citet{Sallum2015} then identified three protoplanets within $\rho$ $\sim$ 0\farcs{}15 ($\approx$ 25 au) (LkCa 15 bcd), one of which was recovered in H$_{\rm \alpha}$ (LkCa 15 b).  Thus, LkCa 15 appeared to show evidence for multiple jovian protoplanets: the first such system ever reported.

%Although SAM is powerful as a method for detecting objects at or below the diffraction limit, 
However, 
%while previous modeling of SAM signals for LkCa 15 assumed the presence of point sources, 
%the presence of bright, spatially extended and morphologically complex disk structure at comparable positions could complicate this fitting and thus the interpretation of the signals.   
the closure phase signals of disks in SAM data can mimic those of protoplanets \citep{Cieza2013, Kraus2013}.  LkCa 15's circumstellar environment as seen in scattered light is complex, including a bright outer dust wall \citep{Thalmann2010,Thalmann2014}.  Additionally, inner dust disk material is now resolved at optical wavelengths and NIR polarimetry out to LkCa 15 bcd-like separations
% similar to LkCa 15 bcd
%, indicating that it might enclose position with these proposed planets
 \citep{Oh2016,Thalmann2016}.    Depending on this dust disk's brightness and spatial extent in (a) total intensity at (b) the longer wavelengths where LkCa 15 bcd were identified (2.2--3.8 $\mu m$), it could instead be the signal masquerading as these protoplanets.   
 %While \textit{direct} imaging data in \textit{total intensity} at 2.2--3.8 $\mu m$ could then decisively determine whether or not these planets exist, 
However, previous 2.2--3.8 $\mu m$ total intensity data lack the image quality/sensitivity to probe these regions \citep{Thalmann2014}.  

In this Letter, we use multi-epoch
%higher Strehl-ratio 
direct imaging observations of LkCa 15 obtained from the Subaru Coronagraphic Extreme Adaptive Optics (SCExAO) project coupled with the Coronagraphic High Angular Resolution Imaging Spectrograph (CHARIS) in the near-infrared
%CHARIS integral field spectrograph
 \citep[$JHK$/1.1--2.4 $\mu m$;][]{Groff2015, Groff2017,Jovanovic2015a} and Keck/NIRC2 in the thermal infrared ($L_{\rm p}$/3.78 $\mu m$).   
These data provide the first direct imaging look at the same wavelengths and in the same locations where previous studies identified the LkCa 15 protoplanets ($K$, $L_{\rm p}$) and thus offer the first decisive test of their existence.    

%Our analysis provides evidence that much, possibly all, of the signal interpreted as LkCa 15 bcd is instead due to extended emission from the inner disk.
%Although LkCa 15 bcd are located at $\sim$ 0.9--1.5 $\lambda$/D in these data, thermal-IR Keck data can nevertheless find planets at contrasts reported for these companions \citep[e.g. $\Delta$L$^\prime$ $\sim$ 7.7 at 0\farcs{}12 or 1.5 $\lambda$/D,][]{Currie2014b}.  After presenting our analysis of these data, we identify required steps to better parse the nature of LkCa 15's inner circumstellar environment and clarify its inventory of planets.

\section{Observations and Data Reduction}
%\subsection{2012 VLT/NaCo Data ($L^\prime$, and [4.05])}
%\subsection{Keck/NIRC2 $L^\prime$ Imaging}
\subsection{SCExAO/CHARIS JHK Direct Imaging/Spectroscopy}
We observed LkCa 15 on UT 2017 September 07 and UT 2018 January 8 using SCExAO coupled with CHARIS operating in low-resolution ($R\sim 20$), 
broadband mode, covering the $JHK$ filters simultaneously ($t_{\rm int}$ =  31 and 19 minutes).    All data were acquired in angular differential imaging mode \citep[ADI;][]{Marois2006}.   
For the September data, given our modest parallactic angle rotation ($\Delta$PA = 60$^{o}$), we also observed a nearby, near-color matched star (V819 Tau) as a contemporaneous point-spread function (PSF) reference\footnote{V819 Tau has a marginal unresolved infrared (IR) excess longwards of 10--15 $\mu m$ \citep{Furlan2009}.   However, we find no hint of a disk in SCExAO/CHARIS data nor in a separate Keck/NIRC2 $L_{\rm p}$ data set. Subaru/HiCIAO $H$-band polarimetry data show that V819 Tau is a non-detection for any disk (J. Hashimoto, pvt. comm).   For our purposes, V819 Tau is effectively a bare stellar photosphere.   
%Thus, there is no evidence for even marginally-resolved disk emission at wavelengths relevant to our study that could possibly impact our results.
}.
 Conditions were excellent (0\farcs{}3--0\farcs{4} $V$-band seeing).  Despite LkCa 15's and V819 Tau's optical faintness ($R$ $\sim$ 11.6, 12.2), we achieved high-quality corrections with a diffraction-limited PSF and the first 8-9 Airy rings  visible.    While we could not directly estimate the Strehl ratio, raw contrasts were similar to those for other stars for which SCExAO's real-time telemetry monitor reported $\approx$ 70\% Strehl in $H$ band.
%: the Strehl ratio estimated from comparing LkCa 15's PSF to that for $\kappa$ And (S.R. $\sim$ 0.9-0.92) was approximately 0.70--0.75 in $H$ band.   
   For the January data, conditions were poorer 
%(S.R (H) $\sim$ 0.55--0.6) 
and we did not observe a PSF reference star, but the parallactic angle motion was larger (120$^{o}$).
%: frame selection identified the highest-quality data.
%   We obtained shorter exposures coupled with frame selection to identify the highest-quality data.

Spectral extraction utilized the cube rectification pipeline from \citet{Brandt2017} and basic image processing was performed as in \citet{Currie2018b,Currie2018a}.   A model spectral energy distribution (SED) reproducing LkCa 15's broadband photometry 
%Appropriate reddened Kurucz model atmospheres ($T_{\rm eff}$ = 4000, log(g) = 4) 
provided spectrophotometric calibration\footnote{LkCa 15 exhibits small-amplitude variability at optical and mid-IR (MIR) wavelengths \citep{Espaillat2011,Rodriguez2017}, with a peak-to-peak value of $\sim$ 0.1 mag.    No clear evidence establishes that LkCa 15 is variable at qualitatively greater level in the $JHK$ bands, let alone at a level that could affect our conclusions.}.   No coronagraphs or satellite spots were used; all stellar PSFs were unsaturated.
% in all channels.     

\subsection{Keck/NIRC2 $L_{\rm p}$ Direct Imaging}
First, we reduced multiple LkCa 15 high-contrast imaging data sets from the Keck Observatory Archive with more than 2$\lambda$/$D$ parallactic angle rotation at LkCa 15 bcd's reported angular separation, selecting  2009 November 21  $L_{\rm p}$ data (PI: L. Hillenbrand; $\Delta$PA = 132$^{o}$.5, $t_{\rm int}$ = 5.4 minutes).  These data have the highest quality of those taken without a coronagraph that may partially occult LkCa 15 bcd and are contemporaneous with the first aperture-masking detection reported in \citet{Kraus2012}.   
Second, we obtained NIRC2 data on 2017 December 9 and 10 for 17.6 and 13.8 minutes with 150$^{o}$ and 160$^{o}$ parallactic motion.   LkCa 15 was observed continuously through transit on the first night; on the second night, we alternated between it and a diskless PSF reference star (V1075 Tau).
All data were acquired in ADI mode using the narrow camera with various dither patterns.   

Keck/NIRC2's adaptive optics (AO) system delivered median Strehl ratios of 0.79 and 
%0.70--0.73
0.77--0.79
 in $L_{\rm p}$ for the 2009 November and two 2017 December data sets, as measured by a modified (for the appropriate pixel scale) observatory-supplied routine \textrm{nirc2strehl.pro}.    
  Stars were unsaturated in all images.   
  %We removed frames with Strehl ratios 2-$\sigma$ below the median value for each set.   
  %For the 2017 data, we observed LkCa 4 and LkCa 14 (xx and xxx minutes) as PSF reference stars. 
 Basic processing followed previous steps used for thermal-IR data with our well-tested broadband imaging pipeline \citep[][]{Currie2011a,Currie2014a},
 including a linearity correction, sky subtraction, distortion correction and bad pixel interpolation, image registration, and flux normalization.  
 % \citep[9.952 mas/pixel;][]{Yelda2010}
  \subsection{PSF Subtraction}
 %\subsection{PSF Subtraction for SCExAO/CHARIS and Keck/NIRC2 Data}
 Because of the complex astrophysical scene within $\rho$ $\sim$ 0\farcs{}5 of LkCa 15, extreme care is needed to properly perform PSF subtraction to avoid misinterpretation
 %and forward-modeling approaches to optimally remove the stellar halo without substantially corrupting astrophysical signals and then interpret the results 
 \citep[][]{Currie2017a}.    For systems like LkCa 15 observed in ADI mode, bright, spatially varying protoplanetary disk emission changes in position angle on the detector over the course of an observing sequence, ``corrupting" the covariance matrices utilized in powerful, widely used least-squares approaches like the \textit{Locally Optimized Combination of Images} (LOCI) and \textit{Karhunen-Lo\'eve Image Projection} (KLIP) algorithms plus successors \citep{Currie2012, Lafreniere2007a,Marois2010b, Marois2014,Soummer2012}.    Additionally, at very small angles, morphological biasing of an astrophysical source in ADI due to self-subtraction can be severe.
 % especially for least-squares-based algorithms.   
 %Set up incorrectly, these algorithms could there preclude, not enable, the detection of planets in disks located at small angles \citep[see case for HD 100546 c;][]{Currie2017b}.
 
Therefore, we adopted the following approach.   First, for data sets obtained with a suitable PSF reference star, we performed reference star differential imaging (RDI) using KLIP and the Adaptive Locally Optimized Combination of Images algorithm \citep[A-LOCI;][]{Currie2012}, where we equate the region used to construct a weighted reference PSF (the optimization zone) and the region over which this PSF is subtracted (the subtraction zone) with the outer radius set to the visible PSF halo, beyond the angles covered by LkCa 15's disk structures ($\rho$ $\approx$ 0\farcs{}75--1\farcs{}1).
%\footnote{In this limiting case, KLIP and A-LOCI are also equivalent.}.    
Second, we performed an ADI-based reduction using A-LOCI on the other data sets using an optimization zone also extending to the PSF halo edge, constructing a weighted reference PSF used to attenuate speckles over smaller annular subtraction zones ($\Delta$r = 2.5--5 pixels).
%\footnote{
%This has been labeled ``conservative LOCI"\citep[e.g.][]{Thalmann2011}.   
%We verified the suitability of this approach for imaging protoplanets at small separations by applying it to archival SPHERE/IRDIS data for PDS 70 and HD 100546 published by \citet{Keppler2018} and \citet{Sissa2018} and easily recovering PDS 70 b,  HD 100546 b, and HD 100546 c \citep{Keppler2018, Quanz2013,Currie2015}.}.   
To further reduce algorithm ``aggressiveness", we applied a rotation gap of $\delta$ $\approx$ 0.5--1 $\lambda$/$D$, while truncating the covariance matrix's diagonal terms with singular value decomposition.   To better suppress residual speckles with the 2018 January CHARIS data, we performed a classical SDI reduction (median-combination of channels rescaled by wavelength) on the ADI/A-LOCI residuals\footnote{CHARIS's large bandpass enables SDI while only partially annealing point sources at LkCa 15 bcd-like separations.}.
%\footnote{Given CHARIS's large bandpass, the range of median movements of a point source at 0\farcs{}1 in a given wavelength slice magnified to align speckles in a slice in $J$, $H$, and $K$ is 0.7--1.1 $\lambda$/D, 0.5--0.6 $\lambda$/D, and 0.5--0.8 $\lambda$/D.   Thus, SDI can operate while only partially annealing any point source at LkCa 15 bcd-like separations.}.

\begin{figure*}[ht]
\centering
\includegraphics[scale=0.185,clip]{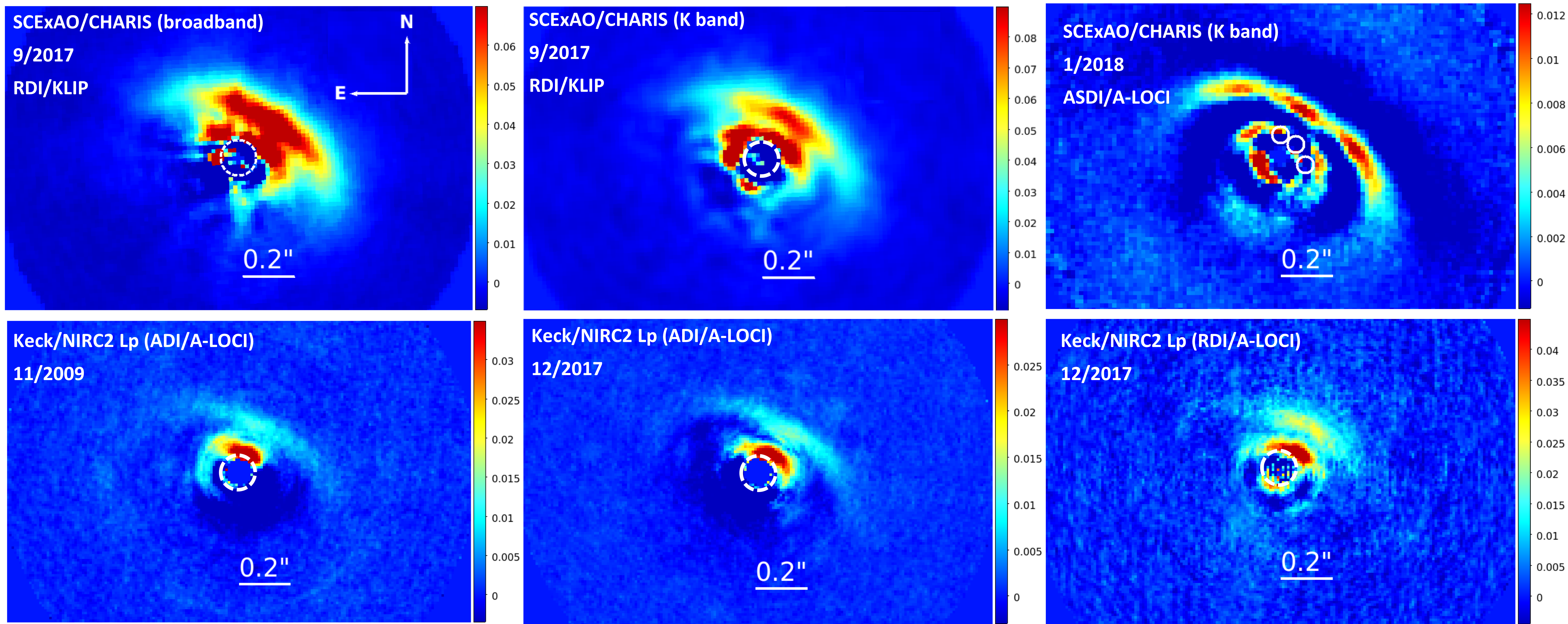}
\vspace{-0.1in}
\caption{LkCa 15 images from SCExAO/CHARIS (top panels) and Keck/NIRC2 (bottom panels).   
The data are processed using different combinations of ADI, SDI, and RDI using the A-LOCI and KLIP algorithms.   All images reveal spatially-extended emission consistent with disk emission, not planets.   A white dashed circle shows a radius of 0\farcs{}06.
% shows the approximate inner working angle.
%The inner working angles (dashed circle in the upper-left panel) are 0\farcs{}05 and 0\farcs{}07 for the CHARIS and NIRC2 data, respectively.
%No image reveals point sources consistent with LkCa 15 bcd; all images show evidence for spatially-extended emission consistent with the inner disk.   
LkCa 15 bcd's positions (circled) from \citet{Sallum2015} trace the edge of the inner disk.   The vertical bars show the intensity scale in units of mJy normalized to one FWHM.}
%; the circle identifies our effective inner working angle of $\theta$ $\approx$ 0\farcs{}06.}
% a least-squares algorithm (A-LOCI) and classical PSF subtraction.}
\label{lkca15images}
\end{figure*}

\section{Detection of the LkCa 15 Inner Dust Disk and Non-detection of LkCa 15 \lowercase{bcd}}
%\section{Detection of Spatially-Extended Emission At Small Angles around LkCa 15 and Non-Detection of LkCa 15 \lowercase{bcd}}
Figure \ref{lkca15images} shows the SCExAO/CHARIS near-IR images in broadband (a median-combination of all channels) and in $K$ band (top panels) and Keck/NIRC2 $L_{\rm p}$ images (bottom panels).
% with various reduction methods.  
All data easily resolve the forward-scattering side of the crescent-shaped outer dust disk wall \citep[e.g.][]{Thalmann2010,Thalmann2014}.
  %The deep January 2018 SCExAO/CHARIS data faintly reveal part of the backside of the disk.     
However, 
no data set reveals direct evidence for LkCa 15 bcd.   Instead, all data resolve another crescent-shaped extended structure interior to the outer disk wall, consistent with the wall of an inner dust disk previously only seen in polarized light \citep{Thalmann2015,Oh2016}.

%We have confirmed that this emission is not an artifact.
   Inspection of individual CHARIS data cubes and NIRC2 images shows that this extended inner disk emission cannot be explained by residual speckle noise that is preserved when images are derotated and combined (for CHARIS and NIRC2) or wavelength-collapsed (for CHARIS).      RDI-reduced images obtained using a range of principal components (for KLIP) or a range of SVD cutoffs (for A-LOCI) all recover the same structure.  For CHARIS, the inner disk is visible in most individual channels, especially those covering the $H$ and $K$ passbands.   Furthermore, ADI and ASDI-reduced images (January 2018 CHARIS data and two of the three Keck/NIRC2 data sets) also show negative self-subtraction footprints of this inner disk\footnote{A separate ASDI reduction of the 2017 September CHARIS data and reduction of other data sets not considered here -- an ADI reduction of archival 2016 October $K_{\rm s}$ SCExAO/HiCIAO data, and ADI reductions of additional archival Keck/NIRC2 $M_{\rm p}$ and $L_{\rm p}$ data from 2012 and 2015 -- likewise show a detection of the inner disk, not planets, albeit with more residual speckle contamination and/or poorer sensitivity.}.  
% or at optical wavelengths.  
%Nov 2009 Kraus two blobs, we clearly show not
%K band shows arc instead of a single point as in Kraus.

We further confirmed that we could have detected LkCa 15 bcd-like planets in absence of disk emission.   To empirically assess our sensitivity to point sources, we injected and attempted to recover model planets with an early L dwarf-like spectrum into our raw LkCa 15 data reduced with RDI (September 2017 CHARIS data and December 2017 NIRC2 data).   We considered the half-field of view opposite the peak brightness of the inner disk and at a range of angular separations\footnote{Typically, contrast curves are derived numerically based on the radial noise profile \citep[e.g.][]{Marois2008a,Currie2011a}.   However, at small angles relevant for this study, corrections to the nominal 3--5 $\sigma$ limits due to finite sample sizes \citep{Mawet2014} are significant.   In particular, the contrast penalty to achieve a Gaussian noise-equivalent 5$\sigma$ limit at 1--2 $\lambda$/D with a False Positive Fraction (FPF) of $\sim$ 2.86$\times$10$^{-7}$ is prohibitively large for a half-field of view \citep[see Figure 6 in ][]{Mawet2014}.  Setting the FPF to 1.35$\times$10$^{-3}$ as recommended by \citet{Mawet2014} for the smallest angles, equivalent to the FPF for a 3$\sigma$ detection in Gaussian statistics, shows that planets with brightnesses comparable LkCa 15 bcd would in fact be recovered at the $>$ 3$\sigma$ level despite residual disk emission.   While residual disk emission at small angles causes the true noise to be overestimated, a substantial positive skew in the noise profile (which itself is uncertain due to finite sample sizes) can cause the FPF to be underestimated \citep{Marois2008a,Currie2014b}. For all these complications, we opt for a more direct, empirical approach of injecting and recovering planets with known contrasts.}.  We varied the brightnesses of these planets with respect to the star to be equal to or fainter than that for LkCa 15 bc at $K$ and $L_{\rm p}$ as reported by \citet{Sallum2015}: $\Delta$$K$ $\sim$ 5.5--6 and $\Delta$$L_{\rm p}$ $\sim$ 5--5.9. 

\begin{figure*}[ht]
\centering
\includegraphics[scale=0.28,clip]{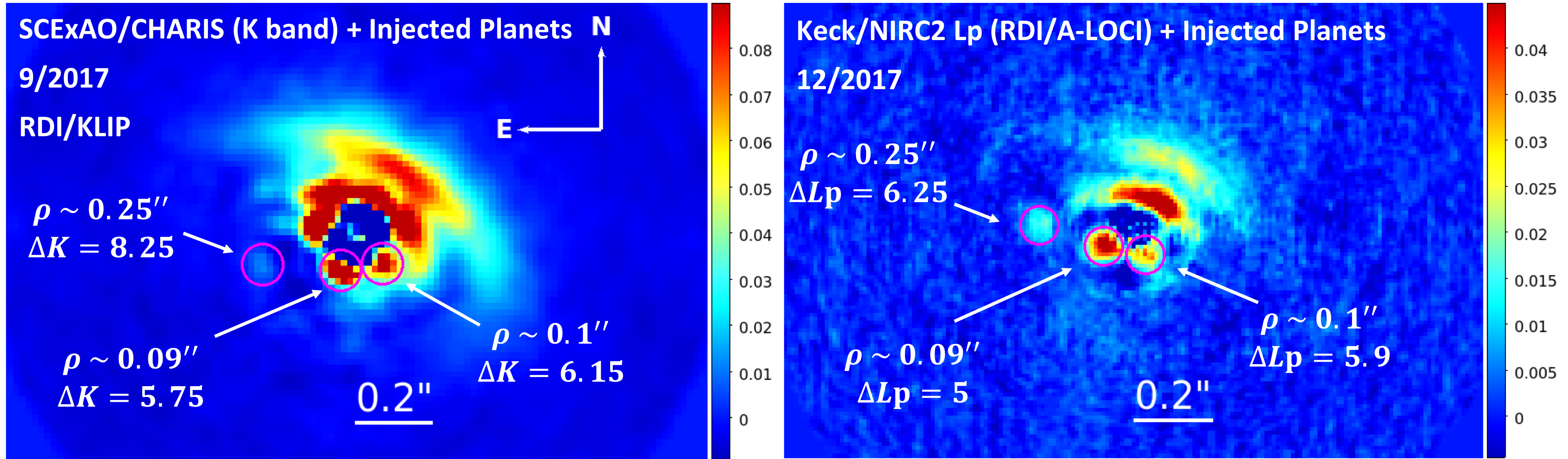}
\vspace{-0.2in}
\caption{September 2017 SCExAO/CHARIS (left) and December 2017 Keck/NIRC2 (right) images with planets (circled) injected into the raw data prior to PSF subtraction with RDI/KLIP and RDI/A-LOCI.   In regions lacking bright extended emission, planets with positions and brightnesses comparable to that reported for LkCa 15 bcd are easily recovered.   Note also that the injected planets are comparable in brightness to the extended emission consistent with an inner disk.   At slightly wider separations ($\rho$ $\sim$ 0\farcs{}25), planets over 1.25--2.5 magnitudes fainter than LkCa 15 bcd are detectable.
%The September 2017 SCExAO/CHARIS image (left) and December 2017 Keck/NIRC2 image (right) with planets injected into the raw data  Comparisons between our observed data (left panels) and forward-models of LkCa 15bcd (middle/right panels) for the September 2017 SCExAO/CHARIS data (top) and November 2009 and December 2017 Keck/NIRC2 data (bottom).   The predicted positions for LkCa 15 bcd in November 2009 and September/December 2017 are $\rho$ $\sim$ 0\farcs{}082, PA=-74$^{o}$ and $\rho$ $\sim$ 0\farcs{}1, PA=-109$^{o}$ for LkCa 15 b; $\rho$ $\sim$ 0\farcs{}083, PA=-6$^{o}$ and $\rho$ $\sim$ 0\farcs{}085, PA=-60$^{o}$ for LkCa 15 c; and $\rho$ $\sim$ 0\farcs{}1, PA=39$^{o}$ and $\rho$ $\sim$ 0\farcs{}08, PA=-1$^{o}$ for LkCa 15 d.
}
\label{lkca15fakeplanet}
\end{figure*}

  Figure \ref{lkca15fakeplanet} shows example RDI-reduced SCExAO/CHARIS and Keck/NIRC2 data sets with injected planets.  The planets' throughputs are high, ranging between 75\% and 100\%.   In spite of some contamination from residual inner disk emission, planets with separations comparable to LkCa 15 bcd ($\rho$ $\sim$ 0\farcs{}09--0\farcs{}1) are detected at LkCa 15 bcd-like contrasts ($\Delta$$K$ $\sim$ 5.75, 6.15; $\Delta$$L_{\rm p}$ $\sim$ 5, 5.9) and visible as point sources.    The contrasts of these recovered planets are similar to limits achieved for diskless stars with SAM in \citet{Kraus2011} and \citet{Lacour2011}.    Planets even fainter than proposed for LkCa 15 bcd -- $\Delta$$K$ $\sim$ 6.5, $\Delta$$L_{\rm p}$ $\sim$ 6.3 -- are detected and identified as point sources at $\rho$ $\sim$ 0\farcs{}09--0\farcs{}1 in regions of the lowest disk emission (not shown).  At wider separations ($\rho$ $\gtrsim$ 0\farcs{}4), our contrast limits are equal to or deeper than  achievable with SAM ($\Delta$$K$, $\Delta$$L_{\rm p}$ $\sim$ 10, 7.5)\footnote{For the 2009 NIRC2 $L_{\rm p}$ data reduced with ADI/A-LOCI, the forward-scattering peak of the inner disk severely self-subtracts point sources injected into the data at $\rho$ $\sim$ 0\farcs{}1: thus, injecting planets into these data as performed for our RDI-reduced data sets substantially underestimates our true sensitivity in absence of a disk.   Nevertheless, planets with LkCa 15 bc-like contrasts are still detectable at LkCa 15 bcd-like separations as well.}.
  %\footnote{For the ADI-reduced .   For the ADI-reduced data, 
%Forward-modeling also confirms that we could have detected individual LkCa 15 bcd-like point sources \textit{in absence of any extended emission}.   For example, we derived a true noise estimate from the half of the image opposite the forward-scattering peaks with fainter extended emission for the RDI/A-LOCI reduced Keck/NIRC2 data.   Despite a severe finite-element correction penalty \citep{Mawet2014},  the 5-$\sigma$ contrast limit is $\Delta$$L_{\rm p}$ $\sim$ 6.3 at 0\farcs{}1, about a factor of 3 (1.5) fainter than the $L_{\rm p}$ contrast for LkCa 15 bc (LkCa 15 d) in \citet{Sallum2015}.  

Comparisons between our images and SAM results strongly suggest that this inner disk emission is the same astrophysical source previously interpreted as the LkCa 15 bcd protoplanets.      
For both CHARIS and NIRC2 data, the inner disk emission extends from $\rho$ $\sim$ 0\farcs{}07 to $\rho$ $\sim$ 0\farcs{}25 ($r_{\rm proj}$ $\approx$ 10--40 au) with an apparent semi-major and semi-minor axis for the emission's peak is $\rho$ $\sim$ 0\farcs{}2 and 0\farcs{}1, respectively ($r_{\rm proj}$ $\approx$ 17--32 au).  In the RDI-reduced data sets, the emission subtends an angle of $\sim$ 100$^{o}$, which is roughly the same position angle range for LkCa 15 bcd reported in \citet{Sallum2015}.    Planet positions reported in \citet{Sallum2015} (circles in the 2018 January CHARIS data) trace %the edge of
 this emission.  The aggregate flux density for LkCa 15 bcd from \citet{Sallum2015} is $\approx$3.7 $\pm$ 1.2 mJy and 5.4 $\pm$ 1.5 in $K_{\rm}$ and $L_{\rm p}$, respectively.    Over the same range of position angles/separations reported for LkCa 15 bcd, the summed inner disk flux densities in the CHARIS $K$ band and NIRC2 $L_{\rm p}$ data reduced using RDI are the same, within uncertainties\footnote{As we found in the immediate preceding analysis, RDI processing induces only modest signal loss for point sources and disks at LkCa 15 bcd-like separations: the throughput-corrected flux density for the inner disk still matches that reported for LkCa 15 bcd combined together.}:  $\approx$2.8 mJy and $\approx$3.9 mJy.  
 %the our photometry and that from \citet{
 
 LkCa 15 images obtained at different wavelengths
% the SCExAO/CHARIS broadband ($JHK$, $\lambda_{\rm o}$ = 1.63 $\mu m$) and $K_{\rm s}$ ($\lambda_{\rm o}$ = 2.18 $\mu$m) images with the NIRC2 $L_{\rm p}$ images 
reveal some evidence for color differences between the spatially resolved inner and outer disk components.   In the SCExAO/CHARIS broadband image ($JHK$, $\lambda_{\rm o}$ = 1.63 $\mu m$), the peak brightness of the inner component is about 30\% higher than the peak of the outer component.  At $K$-band ($\lambda_{\rm o}$ = 2.18 $\mu m$), the peak brightness of the inner disk is about 1.75 times than the outer disk, while at $L_{\rm p}$ the inner disk is more than twice as bright as the outer disk.   The physical origin of these differences will be addressed in \S 4.2.
%Thus, compared to the outer disk, the inner disk appears to be redder in scattered light 

 %The $L^\prime$ direct imaging data, the most direct comparison with published LkCa 15 SAM results \citep{Kraus2012,Sallum2015} directly reveals both the well-studied outer disk and strong signal at angular separations comparable to both the inner disk and the candidate protoplanets for the first time.    
% The outer disk appears as a bright crescent along the minor axis \citep[near side of the disk][]{Thalmann2014}, where the $L_{\rm p}$ images with a localized brightness peak at PA $\approx$ -10$^{o}$.  Although our image appears to faintly trace the far side of the outer disk, the signal-to-noise of this region barely peaks above the background.    The inner disk is spatially resolved, extending about 0\farcs{}13 along the NW minor axis and 0\farcs{}18 along the E major axis.   In the CHARIS image, the entire circumference of the inner disk is resolved; at $L_{\rm p}$,  it appears as a 100$^{o}$  arc of nearly constant brightness, roughly the same range in PA for LkCa 15 bcd reported in \citet{Sallum2015}.  

\begin{figure*}[ht]
\centering
\includegraphics[scale=0.185,clip]{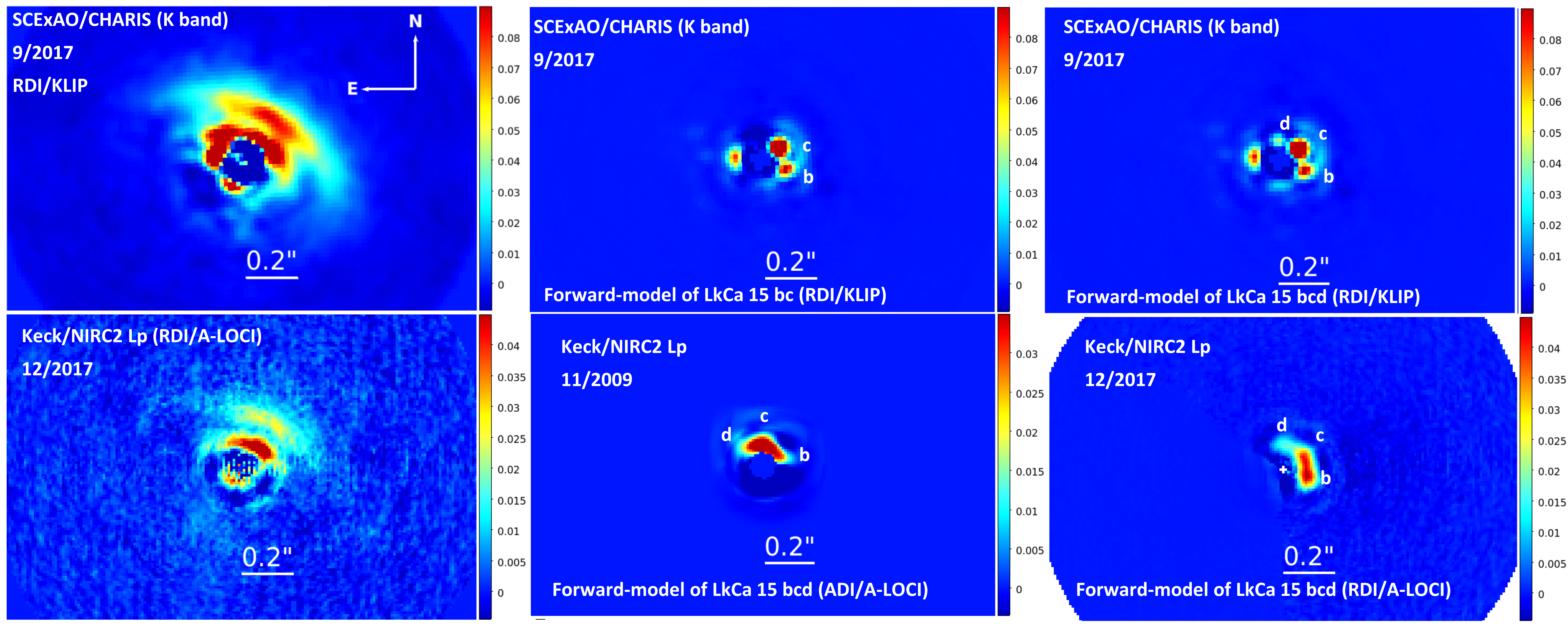}
\vspace{-0.125in}
\caption{Comparisons between our observed data (left panels) and forward-models of LkCa 15bcd (middle/right panels) for the September 2017 SCExAO/CHARIS data (top) and November 2009 and December 2017 Keck/NIRC2 data (bottom).   The predicted positions for LkCa 15 bcd in November 2009 and September/December 2017 are $\rho$ $\sim$ 0\farcs{}082, PA=-74$^{o}$ and $\rho$ $\sim$ 0\farcs{}1, PA=-109$^{o}$ for LkCa 15 b; $\rho$ $\sim$ 0\farcs{}083, PA=-6$^{o}$ and $\rho$ $\sim$ 0\farcs{}085, PA=-60$^{o}$ for LkCa 15 c; and $\rho$ $\sim$ 0\farcs{}1, PA=39$^{o}$ and $\rho$ $\sim$ 0\farcs{}08, PA=-1$^{o}$ for LkCa 15 d.}
\label{lkca15fwdmodplanet}
\end{figure*}

\section{Forward-Modeling of LkCa 15 Images: A Forward-Scattering Inner Dust Disk, Not Multiple Orbiting Planets}
%To more robustly interpret the LkCa 15 images, 
%we compared them to forward-models for LkCa 15 bcd and for LkCa 15's inner disk.
We now compare the LkCa 15 images to forward-models \citep{Marois2010b} for LkCa 15 bcd and an inner disk.  Our analysis
%, which predict the appearance of planets and disks after their signals have been partially annealed by PSF subtraction algorithms.    
adopts the approaches in \citet{Pueyo2016} and \citet{Currie2018a} for KLIP and A-LOCI, using the eigenvalues/eigenvectors in KLIP or coefficients in A-LOCI drawn from the real data and applying them to synthetic planet/disk signals injected into empty data cubes/images.   Our goal is to (1) confirm that the emission we interpret as an inner disk cannot be reproduced by properties previously attributed to LkCa 15 bcd and (2) then explore the general properties of this inner disk. 

 We focused on the
%wavelengths at which LkCa 15 bcd have reported detections and for the 
highest-quality data easily amenable to forward-modeling at wavelengths where LkCa 15 bcd were identified ($K$, $L_{\rm p}$).   Thus, we considered the $K$-band portion of the 2017 September SCExAO/CHARIS data processed with RDI/KLIP, the 2009 November NIRC2 $L_{\rm p}$ data processed with ADI/A-LOCI, and the 2017 December NIRC2 $L_{\rm p}$ data processed with RDI/A-LOCI.
% and $L_{\rm p}$.   filters, where LkCa 1\citet{Kraus2012} and \citet{Sallum2015} report detections of LkCa 15 bcd in the $K_{\rm s}$ and $L_{\rm p}$ filters.   THere Then we compare the real data vs. simulated models .  then we assess whether the models accurately reproduce the data.
%We focus on the K band portion of the CHARIS data and the NIRC2 Lp data.    We use the CHARIS RDI reduction from Sept 2017 and the NIRC2 2009 ADI and 2017 RDI reductions.
\subsection{Planet Forward-Modeling}
We produced forward-models of (a) all three planets (LkCa 15 bcd) and (b) just the two identified in \citealt{Sallum2015} from multiple epochs (LkCa 15 bc), (1) at the planets' last reported positions in \citet{Sallum2015} in November 2014-February 2015, and (2) at the planets' estimated positions in 2009 November, 2017 September, and 2017 December.   To predict the planets' positions in multiple epochs, we adopted the \citeauthor{Sallum2015} astrometry and the \textit{Gaia second data release} (DR2) distance to LkCa 15 (158.9 $pc$), assuming that the planets are on circular orbits in the same plane as the outer disk \citep[$i$ $\sim$ 50$^{o}$, $PA_{\rm minor}$ $\sim$ 60$^{o}$;][]{Thalmann2014,Thalmann2015,Oh2016}.   Their deprojected orbital separations in 2014 November-2015 February are $\sim$ 16--18 au; their position angles change by $\approx$5$^{o}$ yr$^{-1}$ in the orbital plane.   
%In the sky plane, 
%LkCa 15 bcd's motions should change by $\approx$ 0.7, 1.1, and 0.8 $\lambda$/D in the NIRC2 data between November 2009 and December 2017.
%LkCa 15 bcd should be $\approx$ 10--20$^{o}$ (25-35$^{o}$) clockwise (counterclockwise) from their \citeauthor{Sallum2015} positions in September--December 2017 (November 2009): between November 2009 and December 2017 their motions cover $\approx$ 0.7, 1.1, and 0.8 $\lambda$/D.  

We adopted the \citet{Sallum2015} $L_{\rm p}$ photometry for LkCa 15 bcd.   In $K$, we also adopted their LkCa 15 bc photometry.  LkCa 15 d has no claimed detection in $K$ from \citet{Sallum2015}.   We assumed that LkCa 15 d's $K$-$L_{\rm p}$ colors are similar to LkCa 15 bc's and thus adopted $\Delta$$K$ = 7.

\begin{deluxetable*}{lllll}
%\setlength{\tabcolsep}{0pt}
%\tablecaption{Preliminary SCExAO/CHARIS Astrometric Calibration}
\tablecaption{Disk Model Parameters}
\tablewidth{0pt}
%\tablenum{2}
%\tablehead{\colhead{Name} & \colhead{$\chi^{2}_{\nu}$} & \colhead{$\chi^{2}_{\nu, H+K}$}&\colhead{SpT} & \colhead{H$_{\rm cont.}$ index} & \colhead{H$_{2}$K index} & \colhead{Mov. Group} & \colhead{Age} & \colhead{log(L/L$_{\odot}$)}&\colhead{Mass (M$_{\rm J}$)}  & \colhead{References} \\
\tablehead{\colhead{Parameter} & \colhead{Value} & \colhead{} & \colhead{} & \colhead{}}
\tiny
\centering
\startdata
%hline
Global Parameters \\
\hline
Distance & 158.9 $pc$\\
$T_{\rm eff}$   & 4730 K\\
$L_{\star}$   &   1.2 $L_{\odot}$ \\
$R_{\star}$   &   1.65 $R_{\odot}$ \\
$M_{\star}$  &  1.01 $M_{\odot}$ \\
$A_{\rm V}$  & 1.7 \\
  Disk Position Angle ($\theta$) & 60$^{o}$\\
  Dust Size Power Law, $p_{a}$  & 3.5\\
  Dust Carbon Fraction & 0.1\\
\hline
\hline
  Component Parameters & Component 1 & Component 2 & Component 3\\
  \hline
    Disk inclination ($i$) & 50$^{o}$& 51.5$^{o}$& 50$^{o}$\\
  Inner radius, $R_{\rm in}$ (au) & 0.12 & 20 & 55 \\
  Outer radius, $R_{\rm out}$ (au) & 3  & 40& 160 \\
 % Expontential Cutoff Radius, $R_{\rm exp}$ (au) & 1 & - & - \\
  Disk wall radius, $R_{\rm w}$ (au) & 0.12 & 25 & 82.5\\
  Wall shape ($w$) & flat/vertical & rounded/0.3 & rounded/0.25\\
  $M_{\rm dust}$ ($M_{\odot}$ ) & 5$\times$10$^{-8}$  & 7.25$\times$10$^{-6}$  & 1.4$\times$10$^{-3}$ \\
  Radial surface density power law ($\epsilon$) & 1 & 0.5 & 1\\
  Minimum dust size ($a_{\rm min}$, $\mu m$) & 0.1 & 0.6 & 0.1 \\
  Maximum dust size ($a_{\rm max}$, $\mu m$)& 0.25 & 1000 & 1000 \\
  Scale height at inner radius, $H_{\rm o, in}$ & 0.05 & 0.08 & 0.05\\
  Scale height power law, $p_{\rm gas}$ & 1.15 & 1.25 & 1.15\\
\enddata
\vspace{-0.05in}
\tablecomments{The disk component surface density follows $\Sigma$ ($R$ $<$ $R_{\rm w}$) 
$\propto$ 
$R^{-\epsilon}$$\times$exp(-$(\frac{1-R/R_{\rm exp}}{w})^{3}$)
%exp(-($\frac{1-$R$/$R_{\rm exp}$}{$w$}$)$^{3}$) 
and $\Sigma$ ($R$ $\ge$ $R_{\rm w}$) $\propto$ $R^{-\epsilon}$.    The wall shape parameter defines the spatial scale over which the disk surface density increases from $R_{\rm in}$ to $R_{\rm w}$.
See \citet{Mulders2010,Mulders2013} and \citet{Thalmann2014} for detailed explanations of MCMax3D terminology.
}
\label{mcmaxdisk}
\vspace{-0.2in}
\end{deluxetable*}

Figure \ref{lkca15fwdmodplanet} shows forward-models of the LkCa 15 planets for CHARIS $K$-band (top panels) and NIRC2 $L_{\rm p}$ (bottom panels).
% compared to CHARIS and NIRC2 data with intensity contours overlayed.   
%Despite their small reported angular separations, 
 The emission's apparent brightness in the CHARIS data is comparable to the combined brightness proposed for LkCa 15 bcd.   However, 
LkCa 15 bcd would be clearly distinguishable as separate point sources in $K$, whereas the CHARIS data instead show a continuous structure. 
%Their combined brightnesses 
%, after reduced by post-processing,
 %are comparable to the apparent brightness of the arc of emission in the CHARIS data.  
  Thus, the SCExAO/CHARIS data are inconsistent with planets being responsible for this emission.
 
 At $L_{\rm p}$, LkCa 15 bcd's PSFs are partially blended\footnote{
  On the other hand, a forward-model including only LkCa 15 bc, resembling the reconstructed images from 2009 November SAM data \citep{Kraus2012,Sallum2016}, is morphologically inconsistent with our real 2009 November data, as it would reveal the planets as separate point sources.  The SAM image reconstructions in some cases are therefore not faithfully reproducing the spatial distribution of astrophysical signals near LkCa 15.
% On the other hand, a forward-model including only LkCa 15 bc, resembling the reconstructed images from 2009 November SAM data \citep{Kraus2012,Sallum2016}, is morphologically inconsistent with our real 2009 November data.    Like the SCExAO/CHARIS LkCa 15 bc(d) forward-models, this model would reveal the planets as separate point sources: the SAM image reconstructions in some cases are therefore not faithfully reproducing the spatial distribution of astrophysical signals near LkCa 15.
}
%, especially when all three planets are injected, 
%and particularly difficult to distinguish from the visible arc of emission in the 2009 data.   
However, due to orbital motion, the aggregate emission from LkCa 15 bc(d) should rotate clockwise by $\sim$ 35-40$^{o}$ between 2009 and 2017: the emission centroid, measured in the forward-modeled planet images from regions within 50\% of the peak intensity, changes by $\approx$ 1 $\lambda$/$D$.   In contrast, the measured center of mass for this emission in the \textit{real} 2009 and 2017 data is constant to within 0.05--0.1 $\lambda$/$D$, implying a static morphology over 8 yr.  Thus, the Keck/NIRC2 data are inconsistent with planetary orbital motion.   

%Similarly, contrasts derived for V819 Tau in the SCExAO/CHARIS $K$-band data while using LkCa 15 as a PSF reference and 2) V1075 Tau in the Keck/NIRC2 $L_{\rm p}$ data, using LkCa 15 as the PSF reference star with the same algorithm settings.   Despite the relative shallowness of these data sets and  LkCa 15's disk resulting in poorer correlation with the target images, we estimate contrasts of $\Delta$$K$, $L_{\rm p}$ $\sim$ measuring the residuals used LkCa 15 as a PSF reference We computed contrast limits for LkCa 15Using LkCa 15 as a PSF reference star for We derived contrast limits To derive a rough %2017 emission is not rotating, implying a static morphology over an 8-year timescale.  

\begin{figure*}[ht]
\centering
\includegraphics[scale=0.185,clip]{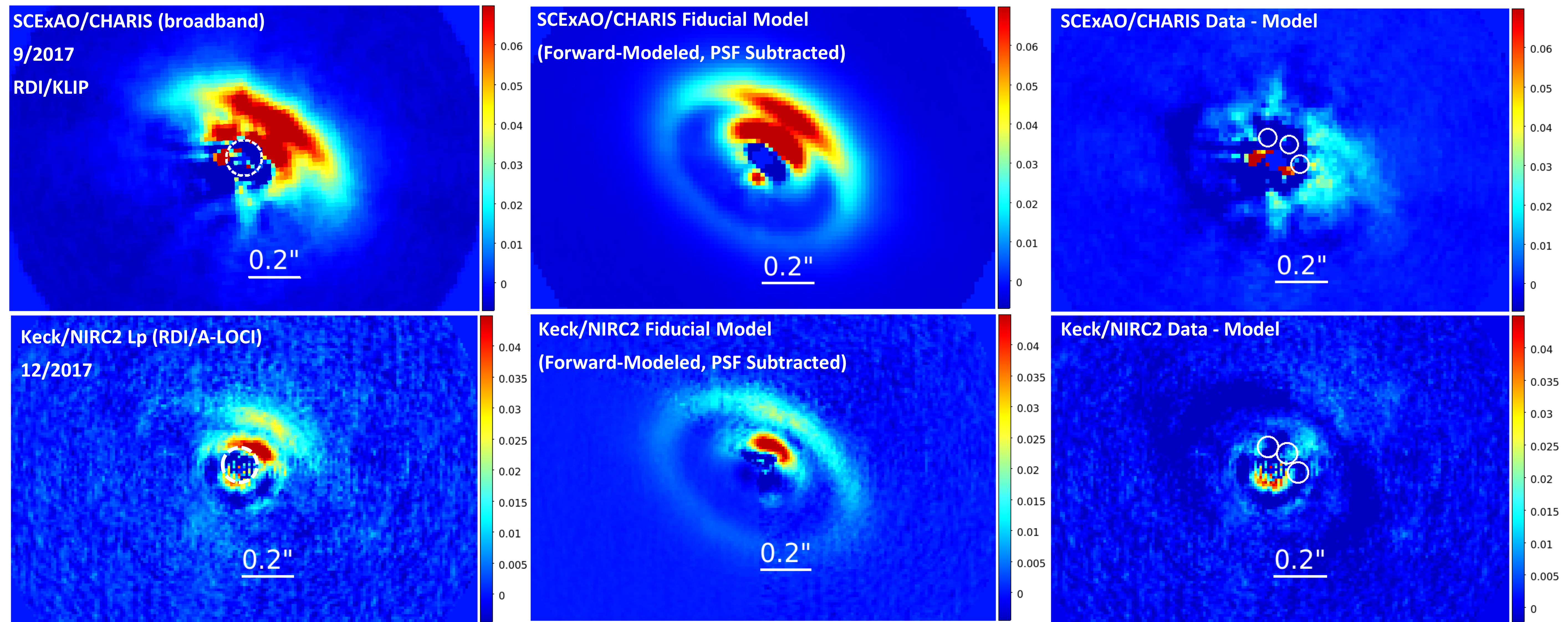}
\vspace{-0.1in}
\caption{Comparing the LkCa 15 SCExAO/CHARIS broadband image (top) and Keck/NIRC2 $L_{\rm p}$ image (bottom) to synthetic disk models.   The left panels show the real data.   The middle panels show the forward-modeled image; the right panel shows the residual image (real data minus model).   The residual image reveals no evidence for embedded planets at LkCa 15 bcd's locations (circles).  The model is produced as-is, not re-scaled in flux to minimize residuals in any dataset.
}
\label{lkca15fwdmoddisk}
\end{figure*}

\subsection{Disk Forward-Modeling}

%We now consider how the emission previously attributed to LkCa 15 bcd is instead total intensity light from a warm inner disk, 
To explore the general properties of inner disk emission previously attributed to LkCa 15 bcd, we
produced and then forward-modeled synthetic scattered-light disk images with SCExAO/CHARIS using the MCMax3D radiative transfer code \citep{Min2009}, adopting the formalism from \citet{Mulders2010,Mulders2013}.
% from \citet{Espaillat2007} and \citet{Mulders2010}.  
Our approach considered three spatially extended components: (1) an optically thick (sub-)au scale hot component responsible for the NIR broadband excess and 10 $\mu m$ silicate feature, (2) a warm component responsible for the inner disk resolved with SCExAO/CHARIS and Keck/NIRC2, and (3) the optically-thick outer disk, which has been resolved in optical/NIR scattered light \citep[e.g.][]{Thalmann2014,Thalmann2016} and with (sub-)millimeter data \citep{Andrews2011,Isella2014}.   Following \citet{Thalmann2014,Thalmann2016}, we envisioned that components 1 and 2 shadow and may be slightly misaligned with the outer disk (component 3).
%, and 3) the optically-thick outer disk, whose directly-illuminated wall is visible in optical/IR scattered light \citep[e.g.][this work]{Thalmann2014,Thalmann2016} and with (sub-)millimeter data \citep{Andrews2011,Isella2014}.   
We explored a small range of component parameters,
% varying the dust grain sizes, masses, radial extents, and scale heights, etc.
 settling on a fiducial model with properties listed in Table 1.    
 %In particular, we a
%We consider three components: 1) an optically thick (sub-)au scale hot component responsible for the near-IR broadband excess at 10 $\mu m$ silicate feature, 2) a 20-30 au-scale warm component responsible for the inner disk we resolve with SCExAO/CHARIS and Keck/NIRC2, and 3) the optically-thick outer disk, whose directly-illuminated wall is visible in optical/IR scattered light \citep[e.g.][this work]{Thalmann2014,Thalmann2016} and with (sub-)millimeter data \citep{Andrews2011,Isella2014}.   We explored a wide range of component properties, varying the dust grain sizes, masses, radial extents, and scale heights, etc.
%\footnote{The main difference between our model and that from \citet{Espaillat2007} and \citet{Mulders2010} is the addition of the 20 au-scale inner disk resolved by SCExAO/CHARIS and Keck/NIRC2.   \citeauthor{Espaillat2007} argued that the 10--20 $\mu m$ portion of the SED require any such dust to be extremely low mass.   We find a far larger dust mass permissible because 1) we consider the shadowing of this component by hot sub-au scale dust (which is optically thick) and 2) the dust is not restricted to submicron sizes.   It produces a scattered light signal comparable to the outer disk because of its larger scale height, allowing it to intercept more starlight and partially shadow the outer disk \citep[see discussion in ][]{Thalmann2015,Thalmann2016}.   Future, focused multi-wavelength disk modeling will identify the best fit set of disk component parameters.}
 %Table 1 lists our adopted model 
 Except for a few \textit{Spitzer}/IRS channels probing the unresolved sub-au component, the model fits LkCa 15's entire SED from the optical to millimeter to within $\sim$ 20-30\% \citep{Andrews2011,Isella2014,Ribas2017}.
%, 2) predicts an inner and outer disk (components 2 and 3) with comparable brightnesses at $K_{\rm s}$ and $L_{\rm p}$, and 3) is consistent with the non-detection of component 2 at 7mm.   
% The chief difference between our model and those from \citet{Espaillat2007} and \citet{Mulders2010} is that 
% In both cases, we start with the inner/outer disk geometries estimated by \citet{Thalmann2014,Thalmann2015,Oh2016}.  
%following recent (forward-)modeling of a similar system HD 163296 in \citet{Rich2018}, we use the Monte Carlo Radiative Transfer code HOCHUNK3D \citep{Whitney2013} to produce synthetic 
%- follow espaillat, mulders, thalmann in disk structure.  opt thick sub au scale disk + outer disk + new component.   
%- rounded wall structure

Figure \ref{lkca15fwdmoddisk} compares the SCExAO/CHARIS broadband image (top) and Keck/NIRC2 $L_{\rm p}$ image with our fiducial model.   The PSF-subtracted model reproduces the brightness and morphology of the inner/outer disk components: the subtraction residuals do not reveal any emission consistent with LkCa 15 bcd.    The peak pixel intensity at positions covering LkCa 15 bcd (circled) is always less than 1/3 (1/4) that predicted for LkCa 15 b(c).   Residuals at $\rho$ $\lesssim$ 0\farcs{}2 that do remain are attributable to slight mismatches with extended disk emission\footnote{For example, weak residuals for NIRC2 just exterior to LkCa 15 c's predicted position correspond to the forward-scattering peak of the outer disk, not the inner disk.   Modified models may better match the combined LkCa 15 data: e.g. faint negative (positive) residuals on the east (west) for the inner/outer disk may be eliminated by introducing pericenter offsets \citep[][]{Thalmann2016}.}.  

While a wide range of models match either the SCExAO/CHARIS or Keck/NIRC2 data, the combined data point toward different grain properties for the three disk components.      A larger minimum dust grain size for the resolved inner disk vs. resolved outer disk ($\sim$ 0.6 $\mu m$ vs. 0.1 $\mu m$) better reproduces the inner disk's redder color and more pronounced forward-scattering peak.  While unresolved, the sub-au disk component requires submicron-sized grains to reproduce the 10 $\mu m$ silicate feature \citep[see also][]{Espaillat2007}.     A future paper will thoroughly analyze LkCa 15's disk structures and derive best-fit parameters.
%; the bottom panels show predicted images at 0.65 $\mu m$ -- similar to $H_{\rm \alpha}$ -- 
%-- comparable to the wavelength where LkCa 15 b was identified from $H_{\rm \alpha}$ imaging -- 
%and 7mm, where \citet{Isella2014} failed to identify disk emission at LkCa 15 bcd-like locations.    
 % The model reproduces the general features of the inner and outer disk.
%The residuals do not reveal any emission that could potentially be LkCa 15 bcd; the peak pixel intensity at $\rho$ $\sim$ 0\farcs{}06--0\farcs{}2 is always less than 1/2 (1/3) that predicted for LkCa 15 b(c).   

Our modeling also (a) implies that LkCa 15's disk structures should be detectable in optical total intensity imaging and (b) is consistent with the millimeter detection of the outer disk and non-detection of the inner disk.   At 0.65 $\mu m$, the inner disk's continuum signal compared to the star (convolved with a gaussian and integrated within 1.5--2 FWHM) near the reported LkCa 15 b position in H$_{\rm \alpha}$
%would range from 5$\times$10$^{-3}$ at the forward-scattering peak near LkCa 15 c to 3$\times$10$^{-3}$ at LkCa 15 b's location (circled):
is just slightly lower than LkCa 15 b's reported H$_{\rm \alpha}$ contrast ($\Delta$F $\sim$ 2.5--5$\times$10$^{-3}$), as is the forward-scattering peak of the outer disk ($\rho$ $\sim$ 0\farcs{}2).   The predicted signal of the forward-scattering peak of the inner disk ($\rho$ $\sim$ 0\farcs{}08) is comparable in contrast to LkCa 15 b ((5--8.5)$\times$10$^{-3}$).    
%In polarized light, the model generally reproduces the structure seen in \citet{Thalmann2015}.
%Thus, the disk should be detectable at $H_{\rm \alpha}$.   
At 7 mm, the model reproduces the outer disk edge's typical intensity, with a characteristic brightness of $\sim$ 24 $\mu$Jy beam$^{-1}$ for a beam size of 0\farcs{}15; for a 0\farcs{}07 beam, it accurately predicts that the inner disk (0.5 $\mu$Jy beam$^{-1}$) would be undetected given a 1$\sigma$ noise floor of 3.6 $\mu$Jy beam$^{-1}$.   

%The fiducial model larg
%When subtracted from the real data, when subtracted from the CHARIS image,  reproduces the general features of the CHARIS data: its sub
 %(top panels) shows the raw best-fit model in $K$ band convolved with the CHARIS PSF, its forward-modeled appearance in CHARIS data, and the subtraction residuals.   The model accurately reproduces both the relative brightness and locations of LkCa 15's disk emission as seen by CHARIS and nulls the signal of the real data.   
 %The model is broadly consistent with multi-wavelength imaging of LkCa 15 (bottom panels), accurately predicting that the inner disk is brighter than the outer disk in optical polarized light but far fainter in the (sub-)millimeter \citep[see][]{Thalmann2016,Andrews2011,Isella2014}.
%, consistent with its non-detection from \citet{Andrews2011} and \citet{Isella2014}.
%the bright polarized optical/$H$ band inner disk signal; its subtraction from the real data yields residuals 
%Figure xxx shows our results comparing the LkCa 15 disk models to the SCExAO/CHARIS data.  

\section{Discussion}
%\citet{Kraus2012} and later \citet{Sallum2015} argue that SAM data for LkCa 15 reveal evidence for up to three individual protoplanets within $\rho$ $\approx$ 0\farcs{}15. 
  Instead of protoplanets, our direct images of LkCa 15 obtained with SCExAO/CHARIS show extended, unresolved inner disk emission.     Forward-modeling shows that the SCExAO data were capable of distinguishing between disk emission and point sources with K band photometry and astrometry reported for LkCa 15's planets by \citet{Sallum2015}.  While \citet{Kraus2012} also identify concentrated emission sources in SAM data, they use a binary (LkCa 15 A+ companions) light distribution model for image reconstruction, which is valid only if the brightness distribution resembles point sources.   Our data show that it does not.   
  
  On the other hand, the inner disk signal is comparable to the total flux density reported for LkCa 15 bcd from \citet{Sallum2015} at $K$ and $L_{\rm p}$.   Thus, we emphasize that the 
  %\citet{Kraus2012} and especially
   \citet{Sallum2015} SAM data likely detected the inner disk at multiple wavelengths.   
     Furthermore, the gaps and misalignments between LkCa 15's resolved disk structures, as well as a warp inferred from the sub-au component \citep{Alencar2018}, may be evidence for unseen jovian planets \citep[][]{DongFung2016}, which   
    could be detected with future facilities \citep[e.g. the \textit{Thirty Meter Telescope};][]{Skidmore2015}.

%  \footnote{\citet{Ireland2013} further argue, based on unpublished $M_{\rm p}$ data, that the emission attributed to LkCa 15 b(cd) is too red to be consistent with scattered starlight.   However, this argument conflates the \textit{individual} signal from each proposed protoplanet at $K_{\rm s}$ and $L_{\rm p}$ with the \textit{aggregate} signal at $M_{\rm p}$.   Taken together, the $K_{\rm s}$ - $M_{\rm p}$ color of ``LkCa 15 bcd" is consistent with scattered starlight.   Our own reduction of these data show a morphology consistent with the inner disk at $L_{\rm p}$}.
  % but rather just interpreted using a model for their morphology now seen to be apparently incorrect.
  % now shown to be .   
  %emission  this is a by-product of using a binaryThe signal from this emission at $K_{\rm s}$ and $L_{\rm p}$ is comparable to that claimed for LkCa 15 bcd.   
  
Our Keck/NIRC2 $L_{\rm p}$ data obtained between 2009 and 2017 reveal this emission to be static.     Based on SAM data taken over a shorter timescale, \citet{Sallum2015,Sallum2016} argued that LkCa 15 bcd astrometry reveals evidence for orbital motion, although different components are detected in different epochs and the combined astrometry appears consistent with stationary sources given large error bars.
%although individual components appear and disappear between epochs and the combined astrometry appears consistent with stationary sources, given large error bars.  
  While the evaluation of our data is straightforward,  several factors may complicate this aspect of SAM data interpretation for LkCa 15.   For example, variable $u-v$ coverage between epochs can induce apparent astrometric offsets when a binary model is assumed in the image reconstruction process (C. Caceres 2019, in preparation).   
  %LkCa 15's inner disk has variable emission \citep{Espaillat2011}, which may induce slight morphological changes and the variable closure phases between epochs.  
   Instead of bare stellar photospheres, the calibrators used for LkCa 15 in \citet{Sallum2015} and especially \citet{Kraus2012} include multiple stars with bright resolved disk emission on the same spatial scale as LkCa 15's disk: some are also highly variable \citep[e.g. GM Aur, UX Tau;][]{Tanii2012, Oh2016b}.    
  
%This work also answers an additional arguments advanced by multiple authors against LkCa 15 bcd being scattered-light disk emission \citep{Kraus2012, Sallum2015,Sallum2016,Ireland2014}.   
Another common argument 
%The most common statement
 is that LkCa 15 bcd are too red to be consistent with scattered-light disk emission
 %, because blue-to-neutral scattered-light colors are typical for protoplanetary dust and LkCa 15 bcd are redder than the star
  \citep{Kraus2012,Ireland2014}.   However, for a system with a pre-transitional disk structure like LkCa 15, (a) scattering can be extremely red because (b) the sub-au dust component contributes significantly to the NIR broadband flux and intercepts (and then re-emits) a significant fraction of the starlight \citep{Mulders2013, Currie2017b}.   The light that LkCa 15's 20 au scale disk ``sees" is then far redder than the star.   Indeed, our fiducial disk model successfully reproduces the brightness of the inner dust disk at $K$ and $L_{\rm p}$.   While \citet{Sallum2015} argued that a disk cannot explain LkCa 15 bc(d) in current SAM data, they use a very simple inclined disk model, not a radiative transfer model.  Additionally, from inspection of their Figure 8, the inner component of this model appears to have semimajor and semiminor axes of $\sim$ 0\farcs{}08 and $\sim$ 0\farcs{}05, which are inconsistent with the larger, spatially resolved and extended disk as resolved at $K$ and $L_{\rm p}$ in this study (0\farcs{}2 and 0\farcs{}1).
  %by SCExAO/CHARISboth the angular separations of be on a spatial scale inconsistent with both twith %structures that likely are on incorrect spatial scales: 
  %an inner (outer) disk component whose semi-minor and semimajor axes appear too small (large) compared to our resolved images.
  %and an outer disk component whose axes are too large.   
 %Furthermore, analysis of the public Keck/NIRC2 aperture masking data from \citet{Kraus2012} shows that the detection of LkCa 15 bcd \textit{can} be reproduced by a disk model (C. Caceres, in prep.).
%that appears to be too small in both is a) simple/not physically motivated with b) a spatial scale that may be inconsistent with that compared to the 
%These statements are incorrect., \citet{Ireland2014} argue that LkCa 15 bcd
%- too red: not true: larger grains + pre-trans structure = red scattering.

%Chiefly, \citet{Kraus2012}, \citet{Ireland2014}, and \citet{Sallum2015} argue that LkCa 15 b(cd) is too red to be consistent with scattered-light disk emission.   
  
 Our analyses do not directly refute the claimed single-epoch MagAO $H_{\rm \alpha}$ detection for LkCa 15 b, which technically remains a candidate companion.  However, they help strengthen arguments voicing strong skepticism.   As LkCa 15 A itself is bright in $H_{\rm \alpha}$ due to accretion its disk structures should have an elevated $H_{\rm \alpha}$ luminosity.   \citet{Mendigutia2018} recently 
  found that LkCa 15's spectroastrometric signature at $H_{\rm \alpha}$ is inconsistent with that of a planet but consistent with a disk.      They rule out $H_{\rm \alpha}$ emission from a LkCa 15 b unless the candidate has an $H_{\rm \alpha}$ contrast fainter than 5.5 mags or a continuum contrast brighter than 6 mags: the $H_{\rm \alpha}$ photometry and continuum upper limits from \citet{Sallum2015} are just barely consistent with these spectroastrometric limits.  Their predicted emitting region for $H_{\alpha}$ is $\rho$ $\sim$ 0\farcs{}07--0\farcs{}16, consistent with our resolved images of LkCa 15's inner disk.
  % than is fainter than its nominal at the contrast at $H_{\rm \alpha}$ ($\Delta$$H_{\rm \alpha}$ = 5.2).
  %showed that $H_{\alpha}$ emission from LkCa 15 could originate in a disk, not a planet.  
  % and in the optical from SPHERE/ZIMPOL \citet{Thalmann2015},
 %   strengthening their conclusion.    
 
 Furthermore, 
 % since LkCa 15 itself is vigorously accreting, the outer, unshadowed regions of the inner disk illuminated by scattered starlight should themselves be bright in $H_{\alpha}$; 
  SPHERE/ZIMPOL data \citep{Thalmann2015} and our modeling show that both the inner disk and outer disk are bright, modest-contrast structures and should be detectable at optical wavelengths covering the MagAO H$_{\alpha}$ observations.   
 %  Yet Extended Data Fig. 2 in \citeauthor{Sallum2015} fails to image the disk with MagAO, suggesting that the $H_{\alpha}$ planet detection may be spurious or it may be a partially subtracted piece of the $H_{\alpha}$-bright disk:   
   Yet \citet{Sallum2015} did not report a disk detection with MagAO, implying that their H$_{\alpha}$ planet detection may instead be spurious or a misidentified, partially subtracted piece of the $H_{\alpha}$-bright disk.  
   %In particular, if the inner disk is eccentric as implied from polarimetry \citep{Thalmann2015,Oh2016}, then 
   % Their data achieved only $\sim$ 2 $\lambda$/D total field rotation, and major axis of the disk could be preferentially preserved even with conservative processing:
    Their quoted position for LkCa 15 b in H$_{\alpha}$ is conspicuously close to the inner disk's major axis.   Given the MagAO observations' poor field rotation (1.5 $\lambda$/$D$ at 0\farcs{}1) and negligibly small rotation gap (5$^{o}$ or $\sim$ 0.12 $\lambda$/D at 0\farcs{}1), any inclined disk at a comparable separation will suffer severe self-subtraction: its residual emission near the major axis would be preferentially preserved and appear point-like.   
    %Absent contemporaneous and prior-epoch aperture masking detections, ``LkCa 15 b" is simply a marginally-significant signal identified in single-epoch whose preferred interpretation is not established.
    %\footnote{Quasi-static speckle noise follows a modified Rician distribution.   With poor field rotation, imperfectly subtracted residual quasi-static speckles with a point source-like spatial scale will not be averaged out in a derotated, sequence-combined image.   As a result, the MagAO false-alarm probabilities, which assume an underlying Gaussian noise distribution, are likely substantially underestimated.}.
    %: the MagAO observations' poor field rotation (1.3 $\lambda$/D at 0\farcs{}1) will induce severe self-subtraction of extended disk emission.
  
     Forward-modeling of both a planet and a disk through the MagAO data -- as performed to assess HD 100546 c \citep{Currie2015} -- could determine which signal better reproduces the images.   However, this test is absent from the \citet{Sallum2015} analysis. 
       % the MagAO $H_{\rm \alpha}$ data are proprietary.
   % we cannot perform this test.
     The MagAO H$_{\rm \alpha}$ data are proprietary, not public, preventing any independent verification that the planet hypothesis is preferred.  
        The public availability of archival Keck/NIRC2 data presented here was crucial in assessing evidence for planets orbiting LkCa 15 from aperture masking.   
        %Future $H_{\rm \alpha}$ imaging with SCExAO/VAMPIRES or VLT/MUSE could help clarify whether the MagAO data identifies an accreting LkCa 15 b \citep{Norris2015, Bacon2004}.
        % clarifying LkCa 15's circumstellar environment and assessing evidence for orbiting planets.
  %\textbf{Thus, we conclude that the claimed near-to-mid IR detections in SAM of LkCa 15 bcd are instead misidentified detections of LkCa 15's inner disk}.  

In summary,  we rule out the proposed LkCa 15 bcd protoplanets as being primarily responsible for emission seen at small angles in SAM data because the emission (a) would be resolved as separate point sources in the SCExAO data (when it is not) and (b) would rotate between 2009 and 2017 Keck/NIRC2 data due to the planets' orbital motion (which it does not).  
Our results also strengthen the argument from \citet{Mendigutia2018} that $H_{\rm \alpha}$ data also likely identifies a disk, not LkCa 15 b.
%While LkCa 15 b remains a candidate due to its claimed $H_{\rm \alpha}$ detection, our results add to the skepticism about this object the argument from \citet{Mendigutia2018} that $H_{\rm \alpha}$ data likewise identifies a disk, not LkCa 15 b, although this object remains a candidate.
%not due to a planet either . 

 Thus, there is currently no clear, direct evidence for multiple protoplanets orbiting LkCa 15.  While the system shows indirect evidence for at least one unseen jovian planet, the bright inner dust disk impedes the detection of this companion(s).  
 %contaminates the signal from any previously-claimed companions at their proposed locations:
  %Previous aperture masking data likely presented the first detections of this inner disk.  
 To confirm jovian companions around LkCa 15 from future observations,
 % Before the direct detection of any planets around LkCa 15 can be decisively established from future observations, 
  the inner disk should be resolved and its effect modeled, removed, and shown to be distinguishable from planets.  Protoplanet candidates identified from similar systems should likewise be clearly distinguished from disk emission through multi-wavelength and/or multi-epoch modeling \citep[e.g.][]{Keppler2018}.  
  
  Distinguishing between disk emission and bona fide protoplanets will continue to be a key challenge for the field of direct imaging \citep[e.g.][this work]{Cieza2013,Kraus2013,Sallum2015a, Ligi2018,Rich2019,Christiaens2019}.

\textbf{Acknowledgements} -- We thank Michiel Min for graciously sharing the MCMax3D code and the Subaru and NASA/Keck Time Allocation Committees for their generous support.   We thank the anonymous referee for a careful, thoughtful review.   Laurent Pueyo, Jun Hashimoto, Christian Thalmann, Catherine Espaillat, Nienke van der Marel, Hannah Jang-Condell, Geoff Bower, and Scott Kenyon provided helpful comments and/or additional, independent assessments of this manuscript.    We emphasize the pivotal cultural role and reverence that the summit of Maunakea has always had within the Hawaiian community.  We are most fortunate to conduct scientific observations from this mountain.  T.C. was supported by a NASA Senior Postdoctoral Fellowship and NASA/Keck grant LK-2663-948181; L.C. was supported by CONICYT-FONDECYT grant No. 1171246.  C.C. acknowledges support from project CONICYT PAI/Concurso Nacional Insercion en la Academia, convocatoria 2015, folio 79150049.   MT is supported by JSPS KAKENHI grant
Nos. 18H05442 and 15H02063.  This work utilized the Keck Observatory Archive (KOA), which is operated by the W. M. Keck Observatory and the NASA Exoplanet Science Institute (NExScI), under contract with the National Aeronautics and Space Administration.   

%Finally, T.C. thanks Mengshu Xu for her exceptional patience and understanding 
 
%\input{model_params.tex}
{}


\begin{thebibliography}{}
%\bibitem[Adams(2010)]{Adams2010}Adams, F. C., 2010, \araa, 48, 47
%\bibitem[Allard et al.(2001)]{Allard2001}Allard, F., Hauschildt, P., Alexander, D. R., et al., 2001, \apj, 556, 357
\bibitem[Alencar et al.(2018)]{Alencar2018}Alencar, S. H. P., Bouvier, J., Donati, J.-F., et al., 2018, A\&A, 620, 195
\bibitem[Andrews et al.(2011)]{Andrews2011}Andrews, S., Rosenfeld, K., Wilner, D. J., Bremer, M., 2011, \apj, 742, L5
%\bibitem[Baraffe et al.(2003)]{Baraffe2003}Baraffe, I., et al., 2003, A\&A, 402, 701
%\bibitem[Augereau et al.(1999)]{Augereau1999}Augereau, J. C., Lagrange, A. M.;, Mouillet, D., 1999, Papaloizou, J. C. B., Grorod, P. A., A\&A, 348, 557
%\bibitem[Avenhaus et al.(2014)]{Avenhaus2014}Avenhaus, H., Quanz, S., Meyer, M. R., et al., 2014, A\&A, 790, 56
%\bibitem[Bacon et al.(2004)]{Bacon2004}Bacon, R., Bauer, S.-M., Bower, R., et al., 2004, \procspie, 5492, 1145
%\bibitem[Barman et al.(2011)]{Barman2011}Barman, T., et al., 2011, \apj\ in press
%\bibitem[Bergfors et al.(2011)]{Bergfors2011}Bergfors, C., et al., 2011, A\&A, 528, 134
%\bibitem[Bailey et al.(2013)]{Bailey2013}Bailey, V., Hinz, P., Currie, T., et al., 2013, \apj, 767, 31
%\bibitem[Baraffe et al.(2003)]{Baraffe2003}Baraffe, I., Chabrier, G., Barman, T. S., et al., 2003, A\&A, 402, 701
%\bibitem[Baraffe et al.(2015)]{Baraffe2015}Baraffe, I., Homeier, D., Allard, F., Chabrier, G., 2015, A\&A, 577, 42
%\bibitem[Beuzit et al.(2008)]{Beuzit2008}Beuzit, J.-L., et al., 2008, SPIE, 7014, 41
%\bibitem[Berriman et al.(1994)]{Berriman1994}Berriman, G. B., Boggess, N. W., Hauser, M. G., et al., 1994, \apj, 431, L63
%\bibitem[Biller et al.(2007)]{Biller2007}Biller, B., et al., 2007, \apjs, 173, 143
%\bibitem[Biller et al.(2010)]{Biller2010}Biller, B., et al., 2010, \apj, 720, 82L
%\bibitem[Boccaletti et al.(2013)]{Boccaletti2013}Boccaletti, A., Pantin, E., Lagrange, A.-M., et al., 2013, A\&A, 560, L20
%\bibitem[Bonnefoy et al.(2011)]{Bonnefoy2011}Bonnefoy, M., et al., 2011, A\&A, 528, 15L
%\bibitem[Booth et al. (2013)]{Booth2012}Booth, M., Kennedy, G., Sibthorpe, B., et al., 2013, \mnras, 428, 1263
%\bibitem[Boss(1997)]{Boss1997}Boss, A., 1997, Science, 276, 1836
%\bibitem[Bowler et al.(2011)]{Bowler2011}Bowler, B., Liu, M. C., Kraus, A., et al., 2011, \apj, 743, 148
\bibitem[Brandt et al.(2017)]{Brandt2017}Brandt, T. D., Rizzo, M., Groff, T., et al., 2017, JATIS, 3, 8002
%\bibitem[Brittain et al.(2014)]{Brittain2014}Brittain, S., Carr, J., Najita, J., 2014, \apj, 791, 136
%\bibitem[Bromley and Kenyon(2011)]{BromleyKenyon2011}Bromley, B., Kenyon, S. J., 2011, \apj\ submitted
%\bibitem[Burgasser et al.(2002)]{Burgasser2002}Burgasser, A., et al., 2002, \apj, 571, 151L
%\bibitem[Burrows et al.(1997)]{Burrows1997}Burrows, A., et al., 1997, \apj, 491, 856
%\bibitem[Burrows et al.(2006)]{Burrows2006}Burrows, A., Sudarsky, D., Hubeny, I., 2006, \apj, 640, 1063
%\bibitem[Carson et al.(2013)]{Carson2013}Carson, J., Thalmann, C., Janson, M., et al., 2013, \apj, 763, L32
%\bibitem[Chauvin et al.(2004)]{Chauvin2004}Chauvin, G., et al., 2004, A\&A, 425, 29L
%\bibitem[Chauvin et al.(2005)]{Chauvin2005}Chauvin, G., et al., 2005, A\&A, 438, 29L
%\bibitem[Chiang et al.(2009)]{Chiang2009}Chiang, E., Kite, E., Kalas, P., et al., 2009, \apj, 693, 734
%\bibitem[Chen et al.(2011)]{Chen2011}Chen, C., Mamajek, E., Bitner, M., et al., 2011, \apj, 738, 122
%\bibitem[Chen et al.(2015)]{Chen2015}Chen, C., Mittal, T., Kuchner, M., et al., 2014, \apjs, 211, 25
\bibitem[Christiaens et al.(2019)]{Christiaens2019}Christiaens, V., Casassus, S., Absil, O., et al., 2019, \mnras\ in press, arxiv:1905.01860
\bibitem[Cieza et al.(2013)]{Cieza2013}Cieza, L., Lacour, S., Schreiber, M., et al., 2013, \apj, 762, L12
%\bibitem[Currie et al.(2008)]{Currie2008}Currie, T., et al., 2008, \apj, 672, 558
%\bibitem[Currie et al.(2009)]{Currie2009}Currie, T., Lada, C. J, et al., 2009, \apj, 698, 1
%\bibitem[Currie et al.(2010)]{Currie2010}Currie, T., Bailey, V., et al., \apj, 721, 177L
\bibitem[Currie et al.(2011)]{Currie2011a}Currie, T., Burrows, A., Itoh, Y., et al., 2011, \apj, 729, 128
%\bibitem[Currie et al.(2011b)]{CurrieLisse2011}Currie, T., Lisse, C. M., Sicilia-Aguilar, A., et al., 2011, \apj, 734, 115
%\bibitem[Currie et al.(2011b)]{Currie2011b}Currie, T., Thalmann, C., et al., 2011, \apj, 736, 33L
\bibitem[Currie et al.(2012)]{Currie2012}Currie, T., Debes, J., Rodigas, T., et al., 2012, \apj, 760, L32
%\bibitem[Currie et al.(2013)]{Currie2013}Currie, T., Burrows, A., Madhusudhan, N., et al., 2013, \apj, 776, 15
%\bibitem[Currie et al.(2014a)]{Currie2014a}Currie, T., Muto, T., Kudo, T., et al., 2014a, \apj, 796, L30
\bibitem[Currie et al.(2014a)]{Currie2014b}Currie, T., Burrows, A., Girard, J., et al., 2014a, \apj, 795, 133
\bibitem[Currie et al.(2014b)]{Currie2014a}Currie, T., Daemgen, Debes, J., et al., 2014b, \apj, 780, L30
%\bibitem[Currie et al.(2014c)]{Currie2014c}Currie, T., Burrows, A., Daemgen, S., 2014, \apj, 787, 104
\bibitem[Currie et al.(2015)]{Currie2015}Currie, T., Cloutier, R., Brittain, S., et al., 2015, \apj, 814, L27
%\bibitem[Currie et al.(2015)]{Currie2015b}Currie, T., Lisse, C., Kuchner, M., et al., 2015, \apj, 807, L7
%\bibitem[Currie et al.(2016)]{Currie2016}Currie, T., Grady, C., Cloutier, R., et al., 2016, \apj, 819, L26
\bibitem[Currie et al.(2017a)]{Currie2017a}Currie, T., Brittain, S., Grady, C., et al., 2017a, RNAAS, 1, 40
\bibitem[Currie et al.(2017b)]{Currie2017b}Currie, T., Guyon, O., Tamura, M., et al., 2017b, \apj, 836, L15
\bibitem[Currie et al.(2018a)]{Currie2018b}Currie, T., Brandt, T. D., Uyama, T., et al., 2018b, \aj, 156, 291
\bibitem[Currie et al.(2018b)]{Currie2018a}Currie, T., Kasdin, N. J., Groff, T., et al., 2018a, \pasp, 130, 044505
%\bibitem[Debes et al.(2008)]{Debes2008}Debes, J. H., Weinberger, A., Schneider, G., 2008, \apj, 673, L191
\bibitem[Dodson-Robinson and Salyk(2011)]{DodsonRobinson2011}Dodson-Robinson, S., Salyk, C., 2011, \apj, 738, 131
\bibitem[Dong and Fung(2017)]{DongFung2016}Dong, R., Fung, J., 2017, \apj, 835, 146
%\bibitem[Dotter et al.(2008)]{Dotter2008}Dotter, A., Chaboyer, B., Jevremovic, D., et al., 2008, \apjs, 178, 89
%\bibitem[Ehrenreich et al.(2010)]{Ehrenreich2010}Ehrenrieich, D., et al., 2010, A\&A, 523, 73 
\bibitem[Espaillat et al.(2007)]{Espaillat2007}Espaillat, C., Calvet, N., D'Alessio, P., et al., 2007, \apj, 670, L135
\bibitem[Espaillat et al.(2011)]{Espaillat2011}Espaillat, C., Furlan, E., D'Alessio, P., et al., 2011, \apj, 728, 49
%\bibitem[Esposito et al.(2014)]{Esposito2014}Esposito, T., Fitzgerald, M., Graham, J., Kalas, P., 2014, \apj, 780, 25
%\bibitem[Esposito et al.(2012)]{Esposito2012}Esposito, S., et al., 2012, A\&A submitted, arXiv:1203.2735
%\bibitem[Fabrycky and Murray-Clay(2010)]{Fabrycky2010}Fabrycky, D., Murray-Clay, R., 2010, 710, 1408
%\bibitem[Fabrycky et al.(2012)]{Fabrycky2012}Fabrycky, D., et al., 2012, \apj\ submitted, arXiv:1202.6328
%\bibitem[Ford and Chiang(2007)]{FordChiang2007}Ford, E., Chiang, E., 2007, \apj, 661, 602 
%\bibitem[Fukagawa et al.(2009)]{Fukagawa2009}Fukagawa, M., et al., 2009, \apj, 696, 1L
\bibitem[Furlan et al. (2009)]{Furlan2009}Furlan, E., Forrest, W. J., Sargent, B. A., et al., 2009, \apj, 706, 1194
%\bibitem[Galicher et al.(2011)]{Galicher2011}Galicher, R., et al., 2011, \apj, 739, 41L
%\bibitem[Galicher et al.(2014)]{Galicher2014}Galicher, R., Rameau, J., Bonnefoy, M., et al., 2014, A\&A, 565, L4
%\bibitem[Goldreich et al.(2004)]{Goldreich2004}Goldreich, P., Lithwick, Y., Sa'ari, R., 2004, ARAA, 42, 549
%\bibitem[Golimowski et al.(2006)]{Golimowski2006}Golimowski, D., et al., 2006, \aj, 131, 3109
%\bibitem[Grady et al.(2001)]{Grady2001}Grady, C., A., Polomski, E. F., Henning, Th., et l., 2001, \aj 122, 3396
%\bibitem[Grady et al.(2005)]{Grady2005}Grady, C. A., Woodgate, B., Heap, S. R., et al., 2005, \apj, 620, 470
\bibitem[Groff et al.(2015)]{Groff2015}Groff, T. D., Kasdin, N. J., Limbach, M., et al., 2015, \procspie, 9605, 96051
\bibitem[Groff et al. (2017)]{Groff2017}Groff, T. D., Chilcote, J., Brandt, T. D., et al., 2017, \procspie, 10400, 1040016
%\bibitem[Hillenbrand et al.(2002)]{Hillenbrand2002}Hillenbrand, L., et al., 2002, \pasp
%\bibitem[Heap et al.(2000)]{Heap2000}Heap, S., et al., 2000, \apj, 539, 435
%\bibitem[Hinz et al.(2006)]{Hinz2006}Hinz, P., et al., 2006, \apj, 653, 1486
%\bibitem[Hinz et al.(2010)]{Hinz2010}Hinz, P., et al., 2010, \apj, 716, 417
%\bibitem[Howard et al.(2010)]{Howard2010}Howard, A., et al., 2010, Science, 330, 653et al
%\bibitem[Hubeny and Burrows(2007)]{HubenyBurrows2007}Hubeny, I., Burrows, A., 2007, \apj, 669, 1248
%\bibitem[Huelamo et al.(2011)]{Huelamo2011}Huelamo, N., et al., 2011, A\&A, 528, 7L
\bibitem[Ireland and Kraus(2014)]{Ireland2014}Ireland, M., Kraus, A., 2014, \textit{Exploring the Formation and Evolution of Planetary Systems}, IAU Symposium, ed. {Booth}, M. and {Matthews}, B.~C. and {Graham}, J.~R., 299, 199
%\bibitem[Jewitt and Luu(1992)]{Jewitt1992}Jewitt, D., Luu, J., 1992, Nature, 362, 730
\bibitem[Jovanovic et al.(2015a)]{Jovanovic2015a}Jovanovic, N., Martinache, F., Guyon, O., et al., 2015, \pasp, 127, 890
%\bibitem[Jura et al.(1998)]{Jura1998}Jura, M., Malkin, M., White, R. J., et al., 1998, \apj, 505, 897
%\bibitem[Kalas and Jewitt(1995)]{KalasJewitt1995}Kalas, P., Jewitt, D., 1995, \aj, 110, 794
%\bibitem[Kalas et al.(2008)]{Kalas2008}Kalas, P., et al., 2008, Science, 322, 1345
%\bibitem[Kalas et al.(2005)]{Kalas2005}Kalas, P., Graham, J. R., Clampin, M., 2005, Nature, 435, 1067
%\bibitem[Kenyon and Bromley(2008)]{KenyonBromley2008}Kenyon, S., Bromley, B., 2008, \apjs, 179, 451
\bibitem[Kenyon et al.(2008)]{Kenyon2008}Kenyon, S. J., Gomez, M., Whitney, B., 2008, Handbook of Star-Forming Regions
\bibitem[Keppler et al.(2018)]{Keppler2018}Keppler, M., Benisty, M., Muller, A., et al., 2018, A\&A, 617, 44
%\bibitem[Konopacky et al.(2011)]{Konopacky2011}Konopacky, Q., et al., 2011, BAAS
%\bibitem[Kratter et al.(2010)]{Kratter2010}Kratter, K., et al., 2010, \apj, 710, 1375
\bibitem[Kraus et al.(2011)]{Kraus2011}Kraus, A., Ireland, M., Martinache, F., Hillenbrandt, L., 2011, \apj, 731, 8
\bibitem[Kraus \& Ireland(2012)]{Kraus2012}Kraus, A., Ireland, M., 2012, \apj, 745, 5
\bibitem[Kraus et al.(2013)]{Kraus2013}Kraus, S., Ireland, M., Sitko, M., et al., 2013, \apj, 768, 80
%\bibitem[Krist et al.(2012)]{Krist2012}Krist, J., Stapelfeldt, K., Bryden, G., Plavchan, P., 2012, \aj, 144, 45
%\bibitem[Kuzuhara et al.(2013)]{Kuzuhara2013}Kuzuhara, M., Tamura, M., Kudo, T., et al., 2013, \apj, 774, 11
\bibitem[Isella et al.(2014)]{Isella2014}Isella, A., Chandler, C. J., Carpenter, J. M., Perez, L. M., Ricci, L., 2014, \apj, 788, 129
\bibitem[Lacour et al.(2011)]{Lacour2011}Lacour, S., Tuthill, P., Amico, P., et al., 2011, A\&A, 532, L72
\bibitem[Lafreni\`ere et al.(2007)]{Lafreniere2007a}Lafreni\'ere, D., Marois, C., Duyon, R., et al., 2007, \apj, 660, 770
%\bibitem[Lafreniere et al.(2007b)]{Lafreniere2007b}Lafreniere, D., et al., 2007b, \apj, 670, 1367
%\bibitem[Lafreniere et al.(2008)]{Lafreniere2008a}Lafreniere, D., et al., 2008, \apj, 689, 153L
%\bibitem[Lafreniere et al.(2008b)]{Lafreniere2008b}Lafreniere, D., et al., 2008b, \apj
%\bibitem[Lafreni\'ere et al.(2009)]{Lafreniere2009}Lafreni\'ere, D., et al., 2009, \apj, 694, 148L
%\bibitem[Lafreniere et al.(2010)]{Lafreniere2010}Lafreniere, D., et al., 2010, \apj, 719, 497
%\bibitem[Lagrange et al.(2009a)]{Lagrange2009}Lagrange, A.-M., et al., 2009a, A\&A, 493, 21L
%\bibitem[Lagrange et al.(2009b)]{Lagrange2009b}Lagrange, A.-M., et al., 2009b, A\&A, 506, 927 
%\bibitem[Lagrange et al.(2010)]{Lagrange2010}Lagrange, A.-M., et al., 2010, Science, 329, 57
%\bibitem[Lagrange et al.(2012)]{Lagrange2012}Lagrange, A.-M., et al., 2012, A\&A,
%\bibitem[Leggett et al.(2010)]{Leggett2010}Leggett, S., et al., 2010, \apj, 710, 1627
%\bibitem[Levison et al.(2008)]{Levison2008}Levison, H., Morbidelli, A., Van Laerhoven, C., et al., 2008, Icarus, 196, 258
\bibitem[Ligi et al.(2018)]{Ligi2018}Ligi, R., Vigan, A., Gratton, R., et al., 2018, \mnras, 437, 1773
%\bibitem[Lisse et al.(2006)]{Lisse2006}Lisse, C. M., VanCleve, J., Adams, A. C., et al., 2006, Science, 313, 635
%\bibitem[Low et al.(2005)]{Low2005}Low, F., Smith, B., Werner, M., et al., 2005, \apj, 631, 1170
%\bibitem[Liu et al.(2010)]{Liu2010}Liu, M., et al., 2010, SPIE, 7736, 53
%\bibitem[Lunine et al.(1989)]{Lunine1989}Lunine, J., et al., 1989, \apj, 338, 314
%\bibitem[Macintosh et al.(2008)]{MacIntosh2008}Macintosh, B., et al., 2008, SPIE, 7015, 31
%\bibitem[Macintosh et al.(2014)]{Macintosh2014}Macintosh, B., Graham, J., Ingraham, P., et al., 2014, PNAS, 111, 35
%\bibitem[Macintosh et al.(2015)]{Macintosh2015}Macintosh, B., Graham, J. R., Ingraham, P., et al., 2015, Science, 350, 64
%\bibitem[Macintosh et al.(2015)]{Macintosh2015}Macintosh, B., Graham, J., Barman, T., et al., 2015, Science, in press
%\bibitem[Madhusudhan et al.(2011)]{Madhusudhan2011}Madhusudhan, N., Burrows, A., Currie, T., 2011, \apj, 737, 34
\bibitem[Marois et al.(2006)]{Marois2006}Marois, C., Lafreni\'ere, D., Duyon, R.,  al., 2006, \apj, 641, 556
\bibitem[Marois et al.(2008a)]{Marois2008a}Marois, C., Lafreni\'ere, D., Macintosh, B., Doyon, R., 2008a, \apj, 673, 647
\bibitem[Marois et al.(2008b)]{Marois2008b}Marois, C., Macintosh, B., Barman, T., et al., 2008b, Science, 322, 1348
\bibitem[Marois et al.(2010a)]{Marois2010a}Marois, C., Zuckerman, B., Konopacky, Q.,  et al., 2010a, Nature, 468, 1080
\bibitem[Marois et al.(2010b)]{Marois2010b}Marois, C., Macintosh, B., Veran, J.-P., 2010b, \procspie, 7736, 52
\bibitem[Marois et al.(2014)]{Marois2014}Marois, C., Correia, C., Galicher, R., et al., 2014, \procspie, 9148, 91480
\bibitem[Mendigutia et al.(2018)]{Mendigutia2018}Mendigutia, I., Oudmaijer, R. D.; Schneider, P. C., et al. 2018, A\&A, 618, L9
%\bibitem[Marois et al.(2010)]{Marois2010}Marois, C., Zuckerman, B., Konopacky, Q., et al., 2010, Nature, 468, 1080
%\bibitem[Martinache and Guyon(2009)]{Martinache2009}Martinache, F., Guyon, O., 2009, SPIE, 7440, 0
\bibitem[Mawet et al. (2014)]{Mawet2014}Mawet, D., Milli, J., Wahhaj, Z., et al., 2014, \apj, 792, 97
%\bibitem[Mazoyer et al.(2014)]{Mazoyer2014}Mazoyer, J., Boccaletti, A., Augereau, J.-C., et al., 2014, A\&A, 569, 29
%\bibitem[Metchev et al.(2009)]{Metchev2009}Metchev, S., Marois, C., Zuckerman, B., 2009, \apj, 705, 204L
%\bibitem[Milli et al.(2012)]{Milli2012}Milli, J., Mouillet, D., Lagrange, A.-M., et al., 2012, A\&A, 545, 111
\bibitem[Min et al.(2009)]{Min2009}Min, M., Dullemond, C. P., Dominik, C., de Koter, A., and Hovenier, J. W., 2009, A\&A, 497, 155
\bibitem[Mulders et al.(2010)]{Mulders2010}Mulders, G., Dominik, C., Min, M., 2010, A\&A, 512, 11
%\bibitem[Mulders et al.(2011)]{Mulders2011}Mulders, G., Waters, L. B. F. M., Dominik, C., et al., 2011, A\&A, 531, 93
\bibitem[Mulders et al.(2013)]{Mulders2013}Mulders, G., Paardekooper, S.-J., Panic, O., et al., 2013, A\&A, 557, 68
%\bibitem[Mustill and Wyatt(2009)]{Mustill2009}Mustill, A., Wyatt, M., \mnras, 399, 1403
%\bibitem[Nesvold and Kuchner(2015)]{Nesvold2015}Nesvold, E., Kuchner, M., 2015, \apj, 798, 83
%\bibitem[Nesvorny et al.(2015)]{Nesvorny2015}Nesvorny, D., 2015, \aj\ in press, arxiv:1504.06021
%\bibitem[Norris et al.(2015)]{Norris2015}Norris, B., Schworer, G., Tuthill, P., et al., 2015, \mnras, 447, 2894
%\bibitem[Oppenheimer et al.(2013)]{Oppenheimer2013}Oppenheimer, R., Baranec, C., Beichman, C., et al., 2013, \apj, 768, 24
\bibitem[Oh et al.(2016a)]{Oh2016}Oh, D., Hashimoto, J., Tamura, M., et al., 2016a, \pasj, 68, L3
\bibitem[Oh et al.(2016b)]{Oh2016b}Oh, D., Hashimoto, J., Carson, J. C., et al., 2016b, \apj, 831, L7
%\bibitem[Pecaut et al.(2012)]{Pecaut2012}Pecaut, M., Mamajek, E.,  Bubar, E., 2012, \apj, 746, 154
%\bibitem[Perrin et al.(2014)]{Perrin2014}Perrin, M., Maire, J., Ingraham, P., et al., 2014, SPIE, 9147, 3
%\bibitem[Pickles et al.(1998)]{Pickles1998}Pickles, A., et al. 1998, \pasp, 110, 749
\bibitem[Pueyo(2016)]{Pueyo2016}Pueyo, L., 2016, \aj, 824, 117
%\bibitem[Quanz et al.(2011)]{Quanz2011}Quanz, S., Schmid, H. M., Geissler, K., et al., 2011, \apj, 738, 26
%\bibitem[Quanz et al.(2013)]{Quanz2013}Quanz, S., Meyer, M. R., Kenworthy, M., et al., 2013, \apj, 766, L1
%\bibitem[Quanz et al.(2015)]{Quanz2015}Quanz, S., Amara, A., Meyer, M. R., et al., 2015, \apj, 807, 64
%\bibitem[Quillen(2006)]{Quillen2006}Quillen, A., 2006, \mnras, 372, L14
%\bibitem[Rameau et al.(2013)]{Rameau2013}Rameau, J., Chauvin, G., Lagrange, A.-M., et al., 2013, \apj, 779, L26
%\bibitem[Rhee et al.(2007)]{Rhee2007}Rhee, J., Song, I., Zuckerman, B., et al., 2007, \apj, 660, 1556
\bibitem[Ribas et al.(2017)]{Ribas2017}Ribas, I., Espaillat, C., Macias, E., et al., 2017, \apj, 849, 63
\bibitem[Rich et al.(2019)]{Rich2019}Rich, E., Wisniewski, J., Currie, T., et al., 2019, \apj, 875, 38
%\bibitem[Rodigas et al.(2014)]{Rodigas2014}Rodigas, T., Malhotra, R., Hinz, P., 2014, \apj, 780, 65
%\bibitem[Rodigas et al.(2015)]{Rodigas2015}Rodigas, T. J., Stark, C., Weinberger, A., et al. 2015, \apj, 798, 96
%\bibitem[Rodigas et al.(2012)]{Rodigas2012}Rodigas, T. J., et al., 2012, \apj\ in press, arXiv:1203.2619
\bibitem[Rodriguez et al.(2017)]{Rodriguez2017}Rodriguez, J., Andsell, M., Oelkers, R., et al., 2017, \apj, 848, 97
%\bibitem[Robitaille et al.(2007)]{Robitaille2007}Robitaille, T., et al., 2007, \apj, 129, 328
\bibitem[Sallum et al.(2015a)]{Sallum2015a}Sallum, S., Eisner, J., Close, L. M., et al., 2015a, \apj, 801, 85
\bibitem[Sallum et al.(2015b)]{Sallum2015}Sallum, S., Follette, K., Eisner, J., et al., 2015b, Nature, 527, 342
\bibitem[Sallum et al.(2016)]{Sallum2016}Sallum, S., Eisner, J., Close, L., et al., 2016, Proc. SPIE, 9907, 99070D
%\bibitem[Service et al.(2016)]{Service2016}Service, M., Lu, J. R., Campbell, R., et al., 2016, \pasp, 128, 5004
%\bibitem[Schneider et al.(2009)]{Schneider2009}Schneider, G., Weinberger, A. J., Becklin, E. E., et al., 2009, \aj, 137, 53
%\bibitem[Schneider et al.(2014)]{Schneider2014}Schneider, G., Grady, C., Hines, D. C., et al., 2014, \aj, 148, 59
\bibitem[Skidmore et al.(2015)]{Skidmore2015}Skidmore, W., TMT International Science Development Teams, Science Advisory Committee, TMT, et al., 2015, RAA, 15, 1945
%\bibitem[Sissa et al.(2018)]{Sissa2018}Sissa, E., Gratton, R., Garufi, A., et al., 2018, A\&A, 619, 160
\bibitem[Soummer et al.(2012)]{Soummer2012}Soummer, R., Pueyo, L., Larkin, J.,  2012, \apj, 755, L28
%\bibitem[Smith and Terrile(1984)]{SmithTerrile1984}Smith, B., Terrile, R., 1984, Science, 226, 1421
%\bibitem[Song et al.(2012)]{Song2012}Song, I., Zuckerman, B., Bessell, M. S., 2012, \aj, 144, 8
%\bibitem[Spiegel and Burrows(2012)]{Spiegel2012}Spiegel, D., Burrows, A., 2012, \apj, 745, 174
%\bibitem[Su et al.(2009)]{Su2009}Su, K., et al., 2009, \apj, 705, 314
%\bibitem[Sudol and Haghighipour(2012)]{Sudol2012}Sudol, J., Haghighighpour, N., 2012, \apj\ submitted, arXiv:1201.0561
\bibitem[Tanii et al.(2012)]{Tanii2012}Tanii, R., Itoh, Y., Kudo, T., et al., 2012, \pasj, 64, 124
%\bibitem[Thalmann et al.(2009)]{Thalmann2009}Thalmann, C., et al., 2009, \apj, 707, 123L
%
%\bibitem[Thalmann et al.(2013)]{Thalmann2013}Thalmann, C., Janson, M., Buenzli, E., et al., 2013, \apj, 763, L29
%\bibitem[Thalmann et al.(2010)]{Thalmann2010}Thalmann, C., Gracy, C. A., Goto, M., et al., 2010, \apj, 718, L87
\bibitem[Thalmann et al.(2010)]{Thalmann2010}Thalmann, C., Grady, C., Goto, M., et al., 2010, \apj, 718, L87
%\bibitem[Thalmann et al.(2011)]{Thalmann2011}Thalmann, C., Janson, M., Buenzli, E., et al., 2011, \apj, 743, L6
\bibitem[Thalmann et al.(2014)]{Thalmann2014}Thalmann, C., Mulders, G. D., Hodapp, K., et al., 2014, A\&A, 566, 5
\bibitem[Thalmann et al.(2015)]{Thalmann2015}Thalmann, C., Mulders, G. D., Janson, M., et al., 2015, A\&A, 808, L41
\bibitem[Thalmann et al.(2016)]{Thalmann2016}Thalmann, C., Janson, M., Garufi, A., et al., 2016, \apj, 828, L17
\bibitem[Tuthill et al.(2006)]{Tuthill2006}Tuthill, P., Lloyd, J., Ireland, M., et al., 2006, Proc. SPIE, 6272, 103
%\bibitem[Yelda et al.(2010)]{Yelda2010}Yelda, S., et al., 2010, \apj, 725, 331
%\bibitem[Vigan et al.(2015)]{Vigan2015}Vigan, A., Gry, C., Salter, G., et al., 2015, \mnras, 454, 129
%\bibitem[Whitney et al.(2013)]{Whitney2013}Whitney, B., et al., 2013, \apjs
%\bibitem[Wyatt(2008)]{Wyatt2008}Wyatt, M. C., 2008, \araa, 46, 339
%\bibitem[van Leewen(2007)]{vanLeewen2007}
%\bibitem[Zhu(2015)]{Zhu2015}Zhu, Z., 2015, \apj, 799, 16
%\bibitem[Zuckerman et al.(2011)]{Zuckerman2011}Zuckerman, B., et al., 2011, \apj, 732, 61
\end{thebibliography}
\end{document}